\documentclass{aastex631}

\shorttitle{AASTeX v6.3.1 Sample article}
\shortauthors{Ashoorioon et al.}
\graphicspath{{./}{figures/}}

\begin{document}
\title{Dark Matter Cosmology with Varying Viscosity: a		Possible Resolution to the $ S_8 $ Tension}
\author[0000-0003-4795-6694]{Amjad Ashoorioon}
\affiliation{School of Physics,	Institute for Research in Fundamental Sciences (IPM)\\ P.O. Box 19395-5531,  Tehran, Iran}
\author[0000-0001-5498-1228]{Zahra Davari}
\affiliation{School of Physics,	Institute for Research in Fundamental Sciences (IPM)\\ P.O. Box 19395-5531,  Tehran, Iran}
\begin{abstract}
We study varying forms of viscous dark matter and try to address the intriguing tensions of the standard model of cosmology with recent cosmological data, including the Hubble and $S_8$ tensions. We note that by assuming the dark matter viscosity depends on the Hubble parameter, dark matter density, or both, one can improve the statistics. Although the models tend to aggravate the Hubble tension a bit, they tend to reduce the $S_8$ tension, even in comparison with the constant viscosity case. Since similar to viscosity massive neutrinos suppress the power spectrum of matter on small length scales, considering them along with the viscous dark matter, we find that the neutrino mass range is tightened.
\end{abstract}
\section{Introduction}\label{sec:int}
New observations on different scales have led us to a more comprehensive understanding of the universe's evolution based on  the standard $\Lambda$CDM model. However, due to the increasing accuracy of observational data, new issues have emerged on the large cosmic scale, for example, $H_0$ (the value of the Hubble parameter observed today) and $\sigma_8$ (the rms fluctuation of density perturbations at 8 h$^{-1}$Mpc scale) tensions referring to disagreement between the predicted values of today’s expansion rate of the universe and matter clustering from CMB data in comparison with their locally determined quantities~\citep{DiValentino:2020zio, Schoneberg:2021qvd,Amon:2022ycy}.\\
These issues lead to considering alternative approaches beyond the $\Lambda$CDM model, such as general relativity modifications or a new description of dark energy (DE)~\citep{Huterer:2017buf,daSilva:2020mvk,Rezazadeh:2022lsf} or different theories of dark matter (DM)~\cite{Vattis:2019efj}. Finding the reason for these apparent discrepancies has become the main driver of cosmological research.\\
One of the main pillars and components in the standard $\Lambda$CDM model is dark matter. The existence of dark matter in our universe is undoubtedly confirmed by numerous kinds of astrophysical observations on a range of length scales, from galaxy rotation curves and gravitational lensing to large scale structure (LSS) and the cosmic microwave background (CMB).
Nevertheless, the physical nature of DM particles is still unclear after decades of research, largely because cosmic observations are only sensitive to the gravitational effects of DM rather than the properties of its particles. What we already know about DM in the $\Lambda$CDM model is that it is responsible for about 85\% of the universe's matter content and is a non-luminous, dark component of matter which must be largely non-relativistic or cold, massive and collisionless. We consider it as perfect and ideal fluid~\citep{Scott:2018adl, Pan-STARRS1:2017jku}. \\
There are many alternatives to cold dark matter, such as cannibal Dark Matter~\citep{Buen-Abad:2018mas}, decaying dark matter~\citep{Davari:2022uwd}, dynamical dark matter, interacting dark matter\citep{Loeb:2010gj, Archidiacono:2019wdp, Arabameri:2023who} and, the interaction between dark energy and dark matter~\citep{Davari:2018wnx}. In these models, attempts have been made to address some of the tensions of the concordance model by resorting to these alternatives.\\
Among the various models to investigate dark matter from a thermodynamic point of view, an interesting proposal is "an exotic fluid with bulk viscosity".
In~\cite{Anand:2017ktp}, it is predicted a viscosity of the order of $10^{-6} H_0 M_P^2$ by the effective field theory of dark matter fluid on large scales. They claimed that this magnitude of viscosity could resolve the discordance between late time large-scale structural observations and Planck's CMB data.\\
These viscosities in the cold dark matter can be engendered in two different ways. The first type of viscosity is generated due to the self-interaction between the dark matter particles, and it is called "fundamental viscosity".
The second type of viscosity is known as "effective viscosity". This kind of viscosity is expected to be generated on large scales as the integrated effect of the back-reaction of small scales non-linearities~\citep{Anand:2017ktp}.\\
During the 1940s and 1950s, the first theory to study bulk viscosity was developed. It is called Eckart's theory~\citep{Eckart:1940te}. But in the late 1960s and 1970s, it became clear that this type of approach suffered from problems of causality and stability~\citep{Muller:1967zza,Israel:1979wp,Poincare1982}. Therefore, by including the second-order deviations from equilibrium, it led to the second-order or M$\ddot{u}$ller-Israel-Stewart theory \citep{MISNER1967,Weinberg:1971mx,Zimdahl:1996ka}. In addition to introducing a new parameter, called relaxation time, the second order theory is a more complicated theory than the first order one. Therefore, the first-order theory has received more attention. We will also develop this work using the Eckart framework.\\
In this paper, we investigate the dynamics of the universe and linear structure formation using a $\Lambda$ viscous dark matter model. The viscosity could be generated by the interactions of the dark matter particles or the large-scale integrated effect of the small scale non-linear gravitational
phenomenon \citep{Blas:2015tla}. Since the universe is not in a steady state but is expanding at an accelerating rate, such sources for viscosity could be time-dependent too. It is also conceivable that the mass of the dark matter particles is determined by a scalar field which does not have a stable vacuum but is determined by the density of the ambient particles. As the universe expands, the density of particles decreases, leading to an increase in the vacuum expectation value of the scalar (and hence the mass of the particle). The energy density of the coupled system of variable-mass particles (``vamps'') redshifts more slowly than that of ordinary matter  \citep{Anderson:1997un, Mandal:2022yym, Das:2023enn}. This could affect the viscosity of dark matter, which is a measure of its resistance to deformation under shear stress. A higher mass would imply a lower velocity dispersion and a higher viscosity, while a lower mass would imply a higher velocity dispersion and a lower viscosity. Therefore, the viscosity of dark matter may depend on the time evolution of the scalar field and the expansion rate of the universe. Hence, contrary  to \citep{Anand:2017ktp}, we assume that the viscosity parameter is not constant and depends on the dark matter energy density or the Hubble parameter, or both of these quantities. We notice that assuming such dependence, the total $\chi^2$ gets reduced. Also,
we can alleviate the $\sigma_8$ tension better. \\
This paper is structured as follows: in section~\ref{sec1},  we first introduce the theoretical basis of this model and derive the main equations governing the evolution at the background (subsection~\ref{sec1-1}) and perturbation (subsection~\ref{sec1-2})  levels for different scenarios. Then we review the key features of the background evolution of the universe.
We modify the public Boltzmann solver \texttt{CLASS}\footnote{\label{myfootnote0}\url{https://github.com/lesgourg/class_public}}(the Cosmic Linear Anisotropy Solving
System)~\citep{Lesgourgues:2011rh} for each type of dissipative mechanisms in the dark matter to calculate the
cosmological evolution and CMB anisotropies.
In section~\ref{sec:mcmc}, we perform the Markov chain Monte Carlo (MCMC) scans by \texttt{MONTEPYTHON-v3}\footnote{\label{myfootnote}\url{https://github.com/baudren/montepython_public}}~\citep{Audren:2012wb,Brinckmann:2018cvx} with a Metropolis-Hasting algorithm using the Planck 2018 CMB data~\citep{Aghanim:2018eyx}, the Pantheon sample of SNIa data~\citep{Pan-STARRS1:2017jku}, an up-to-date collection of Baryon Acoustic Oscillations (BAO) data~\citep{BOSS:2016wmc}, Cosmic Chronometer (CC) data~\citep{Camarena:2018nbr}, an up-to-date compilation of growth rate/redshift space
	distortions data.
We review the basic properties and the cosmological effects of viscous dark matter models for the best value of free parameters in Section~\ref{sec4} according to the results obtained in Section~\ref{sec3}. We perform similar MCMC analyses in Section~\ref{mass} to constrain the neutrino mass. We will discuss cosmic tensions and finally conclude in Section~\ref{con}.
\section{Phenomenology of viscous  dark matter model}\label{sec1}
\subsection{Background level}\label{sec1-1}
In the beginning, we recall the phenomenology of the viscous dark matter models and investigate the effects of viscosity of DM on the solution of Einstein’s equation. In the framework of general relativity, we consider our flat, homogeneous and isotropic universe, including radiation, baryons, the viscous dark matter as non-perfect fluid, and dark energy as a cosmological constant ($\Lambda$)  is described by the FLRW metric
\begin{equation}\label{frw}
ds^2=-dt^2+a^2(t)(dr^2+r^2d\theta^2+r^2\sin^2\theta d\phi^2),
\end{equation}
where $a(t)$ is the scale factor at the cosmic time $t$.\\
We write the energy momentum tensor for non-ideal CDM fluid in the Landau frame, as
\begin{equation}\label{T}
T^{\mu\nu}=\tilde{T}^{\mu\nu}+\Delta T^{\mu\nu},
\end{equation}
where $\tilde{T}^{\mu\nu}$ is the energy–momentum tensor of a perfect fluid, i.e.
\begin{equation}
\tilde{T}^{\mu\nu}=(\rho+p)u^\mu u^\nu+pg^{\mu\nu}.
\end{equation}
Here $ \rho $ is the energy density and $p$ the pressure in the fluid rest frame and $\Delta T^{\mu\nu}$ in
	the first order gradient expansion is a tiny perturbation that corresponds to all dissipative processes (heat flux, anisotropic-stress, and bulk viscosity)~\citep{Weinberg:1971mx} as
	\begin{eqnarray}\label{dt}
	&&\Delta T^{ij}=\Pi_B\Delta^{ij}+\Pi^{ij},\\
	&&\Delta T^{i 0} =\kappa \left[\Delta^{i \alpha}\nabla_\alpha T+T\frac{\partial u_i}{\partial t}\right] ,\nonumber\\
	&&\Delta T^{00}=0,\nonumber
	\end{eqnarray}
	where the matrix $\Delta^{ij}$ projects to the subspace orthogonal to the fluid velocity: $ \Delta^{ij}=g^{ij}+u^i u^j $  and $\Pi_B$ and $\Pi^{ij}$ represent bulk stress and
	shear stress tensor, respectively defined as
	\begin{eqnarray}
	&&\Pi_B=-\xi\nabla_i u^i,\label{eqb}\\ 
	&&\Pi^{ij}=
	-2\eta\left[\frac{1}{2}(\Delta^{i\alpha}\nabla_{\alpha}u^j+\Delta^{j\alpha}\nabla_{\alpha}u^i)-\frac{1}{3}\Delta^{ij}(\nabla_{\alpha}u^\alpha)\right],\nonumber
	\end{eqnarray}	
	where $\eta$ and $\xi$ represent shear and bulk viscosity. As shear viscosity is a directional process, the cosmological principle precludes its existence at large scales in the FLRW metric. On the other hand, $\kappa$ in the middle relation of equation~\ref{dt} shows the temperature conduction coefficient, which is not very relevant to our analysis. Such effects are closely connected to the coupling of the baryonic matter with the photon radiation, which is important either at times closer to recombination, or in the highly nonlinear regime of structure formation at a small scale (e.g., galaxy formation) and when it becomes important for modeling astrophysical processes~\citep{Barbosa:2017ojt}. None  of these regimes where heat conduction would be of relevance will be considered in this work. Hence, only bulk viscosity is a unique process that is allowed to occur on an expansive, homogeneous, and isotropic background.\\
Let us examine the expressions \ref{eqb} in more detail
\begin{equation}
u^{\gamma}_{;\gamma}=\frac{1}{\sqrt{-g}}\partial_\gamma(\sqrt{-g}u^\gamma)=\frac{1}{\sqrt{-g}}\partial_\gamma(\sqrt{-g})u^\gamma+\partial_\gamma u^\gamma,
\end{equation}
where $$\sqrt{-g}=r^2 \sin^2\theta a^3(t),$$ and
$$(\partial_\gamma(\sqrt{-g})u^\gamma=(\partial_0\sqrt{-g})u^0=3r^2 \sin^2\theta a^2(t)\dot{a}(t),$$
which means that $$\frac{1}{\sqrt{-g}}\partial_0(\sqrt{-g})=3\frac{\dot{a}(t)}{a(t)}=3H,$$
which in turn implies
that the bulk viscosity modifies the effective pressure as $p_{\rm eff}=p-3H\xi$ which consists of a sum of two terms, equilibrium pressure $p$ and bulk viscosity pressure, $-3H\xi$.\\
Choosing a reference frame in which the hydrodynamics four-velocity $u_\mu$ is unitary, we can write Equation~\ref{T} for imperfect viscous dark matter fluid as
\begin{equation}
\tilde{T}^{\mu\nu}=(\rho+p_{\rm eff})u^\mu u^\nu+p_{\rm eff}g^{\mu\nu}.
\end{equation}
From the conservation of energy, $\bigtriangledown_\mu T^\mu_\nu=0$,
assuming that no interactions take place among the cosmic fluid components, the continuity
equations take the form,
\begin{eqnarray}
&&\dot{\rho}_r+4H\rho_r=0 \Longrightarrow \rho_r=\rho_{r0} a^{-4},\\
&&\dot{\rho}_b+3H\rho_b=0\Longrightarrow \rho_b=\rho_{b0} a^{-3},\\
&&\dot{\rho}_{\rm vdm}+3H(\rho_{\rm vdm}-3H\xi)=0,\label{con-dm}\\
&&\dot{\rho}_{\Lambda }=0 \qquad\quad\Longrightarrow \rho_\Lambda=\rho_{\Lambda 0}.\label{con-de}
\end{eqnarray}
The dynamics
of the Universe containing these components is given by the Friedmann equation:
\begin{equation}
H^2=(\frac{\dot{a}}{a})^2=\frac{8\pi G}{3}(\rho_r+\rho_b+\rho_{\rm vdm}+\rho_\Lambda).
\end{equation}
Using the definition of the dimensionless energy density
parameter $\Omega_{i,0}=\frac{\rho_{i,0}}{\rho_{\rm cr,0}}$ , where $\rho_{\rm cr,0}=3H_0^2/(8\pi G)$ is the current critical
energy density at $ a=1 $, the Hubble parameter takes the following form:
\begin{equation}
H^2=H_0^2\left(\Omega_{\rm r0}a^{-4}+\Omega_{\rm b0}a^{-3}+y(a)+\Omega_{\rm \Lambda}\right),
\end{equation}
where we define $y(a)=\rho_{\rm vdm}/\rho_{\rm cr,0}$ and set the present density parameter of radiation (photon+relativistic neutrinos) as
$\Omega_{r0}=\Omega_{\gamma0}(1 + 0.2271 N_{\rm eff})$. Here, we put the effective extra relativistic degrees of freedom  3.046 in agreement with the standard model prediction~\citep{Aghanim:2018eyx}.
Applying the Friedman equation, which takes the form of
$\Sigma_i\Omega_i=1$, we can write $\Omega_{\rm \Lambda}=1-\Omega_{\rm r0}-\Omega_{\rm b0}+y(1)$.
Equation~\ref{con-dm} can be rewritten in terms of the scale factor ($d/dt=aHd/da$) and the definition $y(a)$ as follows,
\begin{equation}\label{y}
a\frac{dy(a)}{da}+3y(a)-\frac{\tilde{\xi}}{H_0}\sqrt{\Omega_{\rm r0}a^{-4}+\Omega_{\rm b0}a^{-3}+y(a)+\Omega_{\rm \Lambda}}=0,
\end{equation}
where we have defined the dimensionless parameter $\tilde{\xi}=24\pi G\xi$.
Equation~\ref{y} can be solved numerically by fixing the value $y(a=1)=\Omega_{\rm dm,0}$.  However, we implement the continuity equation~\ref{con-dm} in \texttt{background.c} of the public Boltzmann solver \texttt{CLASS} and use
the shooting method described in~\cite{Audren:2014bca} to compute the  present-day dark matter density.\\
In the rest of the study, we considered four different functional forms for the bulk viscosity coefficient, $\xi(\rho_{\rm vdm},H,\rho_{\rm vdm}H)$:
\begin{itemize}
	\item[i-] Designated as (M1), we consider $\xi$ as constant. This is the model considered by \citep{Anand:2017wsj}.
	\item[ii-] For the second model (M2), where we assume $\xi$ to be proportional to the powers of the energy density of viscous DM, $\left(\frac{\rho_{\rm vdm}}{\rho_{\rm vdm,0}}\right)^n$ and we would obtain the equation~\ref{y} as
\begin{equation}\label{y2}
  a\frac{dy(a)}{da}+3y(a)-\frac{\tilde{\xi}}{H_0}\sqrt{\Omega_{\rm r0}a^{-4}+\Omega_{\rm b0}a^{-3}+y(a)+\Omega_{\rm \Lambda}}\left(\frac{ y(a)}{\Omega_{vdm,0}}\right)^n=0.
\end{equation}
	\item[iii-] For the third model (M3), we consider $\xi$ to be proportional to the powers of the Hubble parameter, $\left(\frac{H}{H_0}\right)^m$ and the equation~\ref{y} is given as
	\begin{equation}\label{y3}
	a\frac{dy(a)}{da}+3y(a)-\frac{\tilde{\xi}}{H_0}\left(\frac{H}{H_0}\right)^{m+1}=0.
	\end{equation}
	\item[iv-] For the fourth model (M4), the coefficient is proportional to $\xi\left(\frac{\rho_{\rm vdm}}{\rho_{\rm vdm,0}}\right)^n\left(\frac{H}{H_0}\right)^m$, and the equation~\ref{y} is given as
	\begin{equation}\label{y4}
	a\frac{dy(a)}{da}+3y(a)-\frac{\tilde{\xi}}{H_0}\left(\frac{y(a)}{\Omega_{\rm vdm,0}}\right)^{n}\left(\frac{H}{H_0}\right)^{m+1}=0.
	\end{equation}
\end{itemize}	
The value assumed for  free parameter, $\xi$, in all of the mentioned models is positive for thermodynamical reasons. The standard cold dark matter model is recovered in the limit $\tilde{\xi}\rightarrow 0$ for all models.
\subsection{Perturbation Level}\label{sec1-2}
In this subsection, we briefly survey the main mathematical form of the linear perturbation theory within the framework of viscous matter dark cosmologies.
We consider the density and pressure in terms of spatially homogeneous and isotropic background fields with small spatially varying perturbations: $\rho(x,t)=\bar{\rho}(t)+\delta\rho(x,t)$ $p(x,t)=\bar{p}(t)+\delta p(x,t)$. 
We assume  perturbations in the variables $\delta\rho,\delta p\ll \bar{\rho}$, $\bar{p}$ up to linear order, and we only focus on the scalar perturbations. We use the normalized density contrast $\delta\equiv\delta\rho/\bar{\rho}$ to expand the fluid dynamic equations into dimensionless measures of these perturbations and the velocity divergence $\theta\equiv\vec{\nabla}\cdot\vec{v}$.\\
Note that pressure and its perturbation are not independent quantities and they are related to density as:
\begin{equation}\label{wcsca}
w=\frac{p}{\rho},\qquad c^2_s=\frac{\delta p}{\delta \rho}, \qquad c^2_{ad}=\frac{\dot{p}}{\dot{\rho}}=w-\frac{\dot{w}}{3{\cal{H}}(1+w)},
\end{equation}
where $w$, $c_s^2$ and $c_{ad}^2$ are the dynamical equation of state, the speed of sound in the medium and the adiabatic sound speed of the viscous DM respectively. Since cold DM is a pressureless ideal fluid, $w = c^2_s = c^2_ {ad} = 0$.
In the following, we write the dynamical equations which govern the evolution of cosmological perturbations in Fourier space  in Synchronous gauge as obtained in ~\citep{Blas:2015tla}, assuming that $\eta=0$, as follows:

	\begin{eqnarray}\label{eq-delta-s}
	&&\dot{\delta}=-3{\cal{H}}(c_s^2-w)\delta-(1+w)(\theta+\frac{\dot{h}}{2}),\\
	&&\dot{\theta}=-{\cal{H}}(1-3c_{ad}^2)\theta+\frac{c_s^2}{1+w}k^2\delta-\nu k^2\theta,\nonumber
	\end{eqnarray}

and in the Conformal Newtonian gauge as

	\begin{eqnarray}\label{eq-delta-n}
	&&\dot{\delta}=-3{\cal{H}}(c_s^2-w)\delta-(1+w)(\theta-3\dot{\phi}),\\
	&&\dot{\theta}=-{\cal{H}}(1-3c_{ad}^2)\theta+\frac{c_s^2}{1+w}k^2\delta+k^2 \psi-\nu k^2\theta.\nonumber
	\end{eqnarray}
Here, $\nu$ is the kinematic viscosity and defined as $\nu=\frac{\xi}{a(1+w)\rho_{\rm vdm}}$ ~\citep{Blas:2015tla,Anand:2017wsj}. 
One can  calculate using the relations~\ref{wcsca}, the equation of state , the sound and adiabatic speeds  as
	\begin{eqnarray}
	&&w=\frac{-3\xi(\rho_{\rm vdm},{\cal{H}},\rho_{\rm vdm}{\cal{H}}){\cal{H}}}{a\rho_{\rm vdm}},\\ 
	&&c^2_s=\frac{-\xi(m+1)\theta}{a \rho_{\rm vdm}\delta}-\frac{3\xi n {\cal{H}}}{a \rho_{\rm vdm}},\\
	&&c^2_{\rm ad}=-\frac{3\xi n {\cal{H}}}{a\rho_{\rm vdm}}-\frac{3\xi (1+m) (\dot{{\cal{H}}}-{\cal{H}}^2)}{a^2\dot{\rho}_{\rm vdm}}.\label{speed}
	\end{eqnarray}
For the constant value of the bulk viscosity coefficient, equations similar to 2-13 to 2-15 by ~\cite{Anand:2017wsj} are obtained. 
We implement the perturbation equations~\ref{eq-delta-s} - \ref{speed} in module \texttt{perturbation.c} of Boltzmann code \texttt{CLASS}.\\
\section{MCMC ANALYSIS}\label{sec:mcmc}
In this section, we present constraints on four viscous dark matter models that we discussed in the previous section. For MCMC analysis we use the Metropolis-Hastings algorithm  of the cosmological sampling package \texttt{MONTEPYTHON-v3}, connected to an altered version of the Boltzmann Solver \texttt{CLASS}.
We use the following datasets to perform the statistical inference:
\begin{itemize}
	\item CMB:  One of the most important observables in
	cosmology due to its well-understood linear physics and sensibility to cosmological parameters is the CMB. Here, we use the CMB likelihoods based on Planck 2018 Legacy, including Planck lensing power spectrum ($40 \leq \textit{l} \leq 400$) (lensing),	high-l Planck power spectra (plik cross-half-mission, $ 30 \leq l \leq 2508 $) (plik), low-$ l $ Planck temperature ($ 2 \leq l \leq 29 $) (lowl ), and low-l  HFI EE polarization ($ 2 \leq l \leq 29 $)(simall )~\citep{Aghanim:2018eyx}.	
    \item  BAO: Another important piece of data for present and future cosmology is the BAO, thanks again to its well-understood linear physics. We use the BAO measurements from  the Baryon Oscillation Spectroscopic Survey Data Release 12 (BAO-fs-BOSS DR12 in $z=0.38$, 0.51 and 0.61)~\citep{BOSS:2016wmc, eBOSS:2020yzd} and eBOSS DR14-Ly-$\alpha$ combined correlations~\citep{deSainteAgathe:2019voe}.\\
	\item SNe Ia: The Pantheon Super Novae (SN) sample consists of magnitude measurements for 1048 SNe Ia with redshifts $0.01 < z < 2.3$~\citep{Pan-STARRS1:2017jku}. Some recent studies have shown that $ H_0 $ tension depends directly on the SN Ia absolute magnitude, $ M_B $~\citep{Efstathiou:2021ocp}. Therefore, here, we consider  the $ M_B $ as a nuisance parameter in our numerical analysis.
	\item Cosmic Chronometers (CC): The CC approach is very powerful in detecting the expansion history of the 	Universe obtained through the measurements of the Hubble parameters.
	We use the $H(z)$ data points from the most recent, accurate estimates and  model-independent data in the redshift range $ 0.07\leq z \leq 1.965 $ based on the relative age of passively evolving galaxies (the differential age technique). These data are uncorrelated with the BAO data points~\citep{Camarena:2018nbr}.
	\item LSS data: To check whether viscous dark matter models lead to a suppression in the matter power spectrum relative to the CDM, it is important to use LSS data, which further strengthens the bounds.	So we use two  different sets of LSS data in order to observe their effect on our fits:\\
		\textbf{1-} KiDS + Viking 450 (KV450) matter power spectrum shape data~\citep{Hildebrandt:2018yau}: This combined analysis of data from the KiloDegree Survey (KiDS) and the VISTA Kilo-Degree Infrared Galaxy Survey (VIKING) includes  photometric redshift measurements with cosmic shear/weak-lensing observations to measure the matter power spectrum over a wide range of $k$-scales at redshifts between 0.1 and 1.2.\\
		\textbf{2-} Planck SZ (2013):  Another second independent LSS dataset is the Planck SZ which studies the properties of galaxy clusters by measuring the Sunyaev-Zeldovich effect. But it should be noted the measurements of galaxy distribution from the SZ effect depend on a mass bias factor $ (1-b) $ that relates the observed SZ signal to the true mass of galaxy clusters. In Planck SZ (2013), a numerical simulation of the $ S_8 $ measurement is reported by fixing the mass bias to its central value $(1-b) = 0.8$.
		Later,  the Planck SZ (2015) report allowed $ (1-b) $ to vary with a Gaussian prior centered at 0.79. The central value of the resulting $ S_8^{\rm SZ}$ becomes smaller but has a much larger uncertainty,
		$ S_8^{\rm SZ}= 0.744\pm0.034$ and less tension to CMB	measurements~\citep{Planck:2015lwi}. However, since the central value $\sigma_8$ of the SZ (2013) analysis is consistent with many low-redshift measurements, we chose this data set for our analysis~\citep{Zu:2023rmc}.\\
		\textbf{3-}WiggleZ $P(k)$ data: Since dark energy has an effect on the expansion history of the Universe and on the growth of cosmological structures, we also use WiggleZ data in this study.
		The WiggleZ Dark Energy Survey is a survey	to measure the large scale structure of the Universe by mapping the distance-redshift relation with baryon acoustic oscillations~\citep{Kazin:2014qga}.
\end{itemize}
We employ Bayes's theorem to find the posterior  probability distribution of a parameter, $\theta$, given a model, $M$, and a dataset, $D$, is given by
	\begin{equation}
	P(\theta|D,M)=\frac{{\cal{L}}(\theta)\pi(\theta)}{E},
	\end{equation}
where $\pi(\theta)$ (or $P(\theta|M)$) is the prior parameter distribution, ${\cal{L}}$ (or $P(D|\theta, M )$) is the likelihood, and the normalizing coefficient $E$(or $P(D|M)$ ) is the model likelihood or evidence, and can be used to evaluate the probability of the model as a whole. For the parameter priors, we assume a uniform probability distribution~\citep{Parkinson:2012vd}. We implement all the likelihoods from the public cosmological MCMC package \texttt{Monte Python} and the total $\chi^2$ ($\equiv-2\ln {\cal{L}}$) is calculated as
	\begin{equation}
	\chi^2_{{}_{\rm tot}}=\chi^2_{{}_{\rm CMB}}+\chi^2_{{}_{\rm BAO}}+\chi^2_{{}_{\rm SNIa}}+\chi^2_{{}_{\rm CC}}+\chi^2_{{}_{\rm KV450}}+\chi^2_{{}_{\rm SZ}}+\chi^2_{{}_{\rm WiggleZ}}.
	\end{equation}
\begin{figure}
	\centering
	\includegraphics[width=12cm]{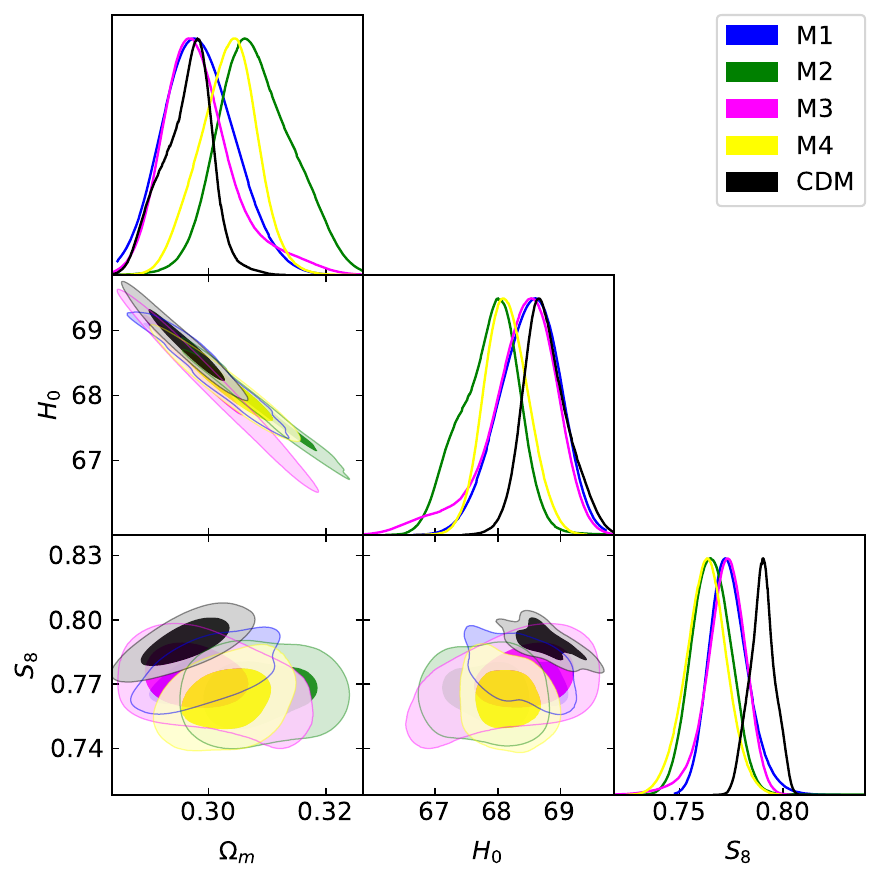}
	
	\caption{68\% and 95\% constraint contours on the matter density parameter $ \Omega_m $ , the Hubble parameter, $H_0$ and $S_8$. It is obtained by considering the Planck+BAO+CC+SNIa+LSS data
		\label{fig1}}
\end{figure}
As a comparison, we performed a similar MCMC scan for the CDM model, also based on the same likelihood.
Here, in addition to the six free parameters of the standard minimum model, i.e. $(\Omega_b,\Omega_{DM},100\theta_{\rm MC},\ln 10^{10}A_s,n_s,\tau_{{}_{\rm reio}})$, the viscous dark matter models introduced in the previous section include at most $(\xi, n, m)$ in the M4 model. Model M1 has one parameter and models M2 and M3 have two parameters. The convergence of chains for each parameter is measured by the Gelman-Rubin criterion and one can obtain acceptable $ R-1 $ values (i.e. below 0.01 for every parameter) with an iterative strategy~\citep{Gelman:1992zz} and the average acceptance rate is around 0.2.\\
In Table~\ref{tab:best}, we report the best and the mean values and 68\% CL intervals for main parameters including the total matter density
parameter ($\Omega_m=\Omega_B+\Omega_{DM}$), the present-day
expansion rate of the Universe or the Hubble constant, $H_0$, and $S_8=\sigma_8\sqrt{\Omega_m/0.3}$ in different scenarios for two MCMC analyzes. We also show posterior distributions (1$\sigma$ and 2$\sigma$ intervals) as dark and light shaded contours for  MCMC analysis in Figure~\ref{fig1}. Some points in this plot need to be stressed. First, we can see that the decrease in $ H_0 $ values is associated with the increase of $ \Omega_m $ values and vice versa, in both viscous and cold dark matter scenarios, but the behavior of  $\Omega_m-S_8$ and $H_0-S_8$ in two scenarios are different. Second, it seems that assuming viscosity improves the $ S_8 $ tension for the viscous dark matter scenario. We will discuss this further in  section~\ref{sec:s8}.\\
\begin{table}
	\centering
	\caption{The mean and best-fit values
		of the free parameters obtained
		for different viscous dark matter scenarios and the CDM model by considering  different data sets. It is obvious
		from these values that the $ H_0 $ tension is not solved, since $S_8$ tension is unraveled.}
	\begin{tabular}{|lll|}
		\hline
		\multicolumn{3}{|c|}{M1}               \\ \hline
		\multicolumn{1}{|l|}{Param} & \multicolumn{1}{l|}{best-fit} & $mean\pm\sigma$ \\ \hline
		\multicolumn{1}{|l|}{$\Omega_m$}    & \multicolumn{1}{l|}{$0.299$}     &   $0.299^{+0.005}_{-0.007}$     \\ \hline
		\multicolumn{1}{|l|}{$H_0$}     & \multicolumn{1}{l|}{$ 68.00 $}     &  $68.50^{+0.51}_{-0.43}$    \\ \hline
		\multicolumn{1}{|l|}{$S_8$}    & \multicolumn{1}{l|}{$0.777$}     & $0.774^{+0.008}_{-0.010}$      \\ \hline
		\multicolumn{1}{|l|}{$\xi(10^{-5})$}    & \multicolumn{1}{l|}{$0.031$}     &   $0.026^{+0.005}_{-0.004}$   \\ \hline
			\multicolumn{3}{|c|}{M2}               \\ \hline
		\multicolumn{1}{|l|}{Param} & \multicolumn{1}{l|}{best-fit} & $mean\pm\sigma$ \\ \hline
		\multicolumn{1}{|l|}{$\Omega_m$}    & \multicolumn{1}{l|}{$0.307$}     &  $0.308^{+.006}_{-0.007}$    \\ \hline
		\multicolumn{1}{|l|}{$H_0$}     & \multicolumn{1}{l|}{$68.00$}     &  $67.87^{+0.52}_{-0.40}$    \\ \hline
		\multicolumn{1}{|l|}{$S_8$}    & \multicolumn{1}{l|}{$0.757$}     &  $0.765\pm0.01$    \\ \hline
		\multicolumn{1}{|l|}{$\xi(10^{-5})$}    & \multicolumn{1}{l|}{$0.041$}     &   $0.044^{+0.013}_{-0.020}$   \\ \hline
		\multicolumn{1}{|l|}{$n$}    & \multicolumn{1}{l|}{$-0.038$}     &  $-0.031^{+0.009}_{-0.014}$    \\ \hline
			\multicolumn{3}{|c|}{M3}               \\ \hline
		\multicolumn{1}{|l|}{Param} & \multicolumn{1}{l|}{best-fit} & $mean\pm\sigma$ \\ \hline
			\multicolumn{1}{|l|}{$\Omega_m$}    & \multicolumn{1}{l|}{$0.310$}     &  $0.299^{+.004}_{-0.007}$    \\ \hline
		\multicolumn{1}{|l|}{$H_0$}     & \multicolumn{1}{l|}{$68.64$}     &  $68.37^{+0.63}_{-0.39}$    \\ \hline
		\multicolumn{1}{|l|}{$S_8$}    & \multicolumn{1}{l|}{$0.795$}     &  $0.772^{+0.010}_{-0.008}$    \\ \hline
		\multicolumn{1}{|l|}{$\xi(10^{-5})$}    & \multicolumn{1}{l|}{$0.112$}     &   $0.036^{+0.014}_{-0.025}$   \\ \hline
		\multicolumn{1}{|l|}{$m$}    & \multicolumn{1}{l|}{$0.044$}     &  $0.004^{+0.031}_{-0.037}$    \\ \hline
			\multicolumn{3}{|c|}{M4}               \\ \hline
		\multicolumn{1}{|l|}{Param} & \multicolumn{1}{l|}{best-fit} & $mean\pm\sigma$ \\ \hline
				\multicolumn{1}{|l|}{$\Omega_m$}    & \multicolumn{1}{l|}{$0.308$}     &  $0.304\pm0.005$    \\ \hline
	\multicolumn{1}{|l|}{$H_0$}     & \multicolumn{1}{l|}{$67.91$}     &  $68.14^{+0.33}_{-0.37}$    \\ \hline
	\multicolumn{1}{|l|}{$S_8$}    & \multicolumn{1}{l|}{$0.768$}     &  $0.763\pm0.010$    \\ \hline
	\multicolumn{1}{|l|}{$\xi(10^{-5})$}    & \multicolumn{1}{l|}{$0.002$}     &   $<0.011$   \\ \hline
	\multicolumn{1}{|l|}{$n$}    & \multicolumn{1}{l|}{$-0.032$}     &  $-0.036^{+0.011}_{-0.017}$    \\ \hline
			\multicolumn{1}{|l|}{$m$}    & \multicolumn{1}{l|}{$-0.038$}     &  $-0.037^{+0.013}_{-0.020}$    \\ \hline
			\multicolumn{3}{|c|}{cdm}               \\ \hline
		\multicolumn{1}{|l|}{Param} & \multicolumn{1}{l|}{best-fit} & $mean\pm\sigma$ \\ \hline
		\multicolumn{1}{|l|}{$\Omega_m$}    & \multicolumn{1}{l|}{$0.299$}     &  $0.295^{+0.005}{-0.004}$    \\ \hline
		\multicolumn{1}{|l|}{$H_0$}     & \multicolumn{1}{l|}{$68.64$}     & $68.76^{+0.29}_{-0.39}$     \\ \hline
		\multicolumn{1}{|l|}{$S_8$}    & \multicolumn{1}{l|}{$0.794$}     & $0.790\pm0.006$     \\ \hline
	\end{tabular}\label{tab:best}
\end{table}
Next, to check whether the fit is good, and also to choose the best and most compatible model with the observational data, we use the simplest method that is usually used in cosmology, which is called the least squares method,  $\chi^2_{{}_{\rm tot}}$. In this case, the model with smaller $\chi^2_{{}_{\rm tot}}$ is taken to be a better fit to the data~\citep{Davari:2021mge}. Comparing different viscous models, we note that all models of viscose dark matter do better than CDM, even for the M1 scenario, where the viscosity parameter is assumed to be constant. The $ \chi^2_{{}_{\rm tot}}$ of the M4 model for the second MCMC analysis is much smaller than the other cases, showing an improvement of about 18 units of  $\chi^2_{{}_{\rm tot}}$ relative to the CDM model, while only adding three new parameters to the fit. 
However, one can have the impression that the model with the lowest  $\chi^2_{{}_{\rm tot}}$ is not necessarily the best model because adding more flexibility with extra parameters will normally lead to a lower  $\chi^2_{{}_{\rm tot}}$.That means that a poorly parameterized model should be penalized. In this work, the M1 model has one parameter, M2 and M3 have two parameters and M4 has three parameters more than the CDM scenario.  In order to deal with model selection, a standard approach is to compute the Akaike Information Criterion (AIC),
\begin{equation}\label{aic}
AIC=\chi^2_{\rm min}+2M+\frac{2M(M+1)}{N-M+1}.
\end{equation}
Above $M$ is the number of free parameters in the model and $N$ is the number of data points. Therefore, in this criterion, we reduce the practical influence of $M$ in $\chi^2_{\rm min}$ in favor of a model with a lower free parameter. Since $N\geq M$ we neglect the third term in the Equation~\ref{aic}. We report the result of MCMC analysis for the best-fit $\chi^2$ for observational Planck and total data sets and compare the viscous DM models to the reference CDM model in Table~\ref{tab:comp}.  The results of this analysis can be interpreted with the Jeffreys' scale as follows: adding time-dependence to the bulk viscosity of dark matter, one can improve the statistics. In particular, adding one, two, and three new parameters to the fit in the viscous dark matter scenarios, the $AIC$ for them is much smaller than the CDM model. It shows an improvement of about 11.25, 9.67, 9.42 and 5.518 units of AIC   relative to the CDM model for M4, M3, M2 and M1 scenarios, respectively.\\
\begin{table}
		\centering
	\caption{The result of MCMC analysis for all reviewed models in this study for the best-fit $\chi^2$.}
	\begin{tabular}{|llllll|}
		\hline
		\multicolumn{6}{|c|}{Total data}                                                                                                                                   \\ \hline
		\multicolumn{1}{|l|}{}                           & \multicolumn{1}{l|}{M1} & \multicolumn{1}{l|}{M2} & \multicolumn{1}{l|}{M3} & \multicolumn{1}{l|}{M4} & CDM \\ \hline
				\multicolumn{1}{|l|}{$\chi^2_{\rm min}$}         & \multicolumn{1}{l|}{$ 4848.68 $}   & \multicolumn{1}{l|}{$ 4842.45 $}   & \multicolumn{1}{l|}{$ 4842.20 $}   & \multicolumn{1}{l|}{$ 4838.32 $}   &  $4855.87  $  \\ \hline
		\multicolumn{1}{|l|}{$\rm AIC-AIC_{\rm CDM} $}                           & \multicolumn{1}{l|}{$ -5.518 $}   & \multicolumn{1}{l|}{$ -9.42 $}   & \multicolumn{1}{l|}{$ -9.67 $}   & \multicolumn{1}{l|}{$ -11.55 $}   &  $ - $   \\ \hline
	\end{tabular}\label{tab:comp}
\end{table}	
\section{Observtional Effects of Viscous DM Models}\label{sec3}
\subsection{Background level}
\begin{figure}
\gridline{	\includegraphics[width=0.5\linewidth]{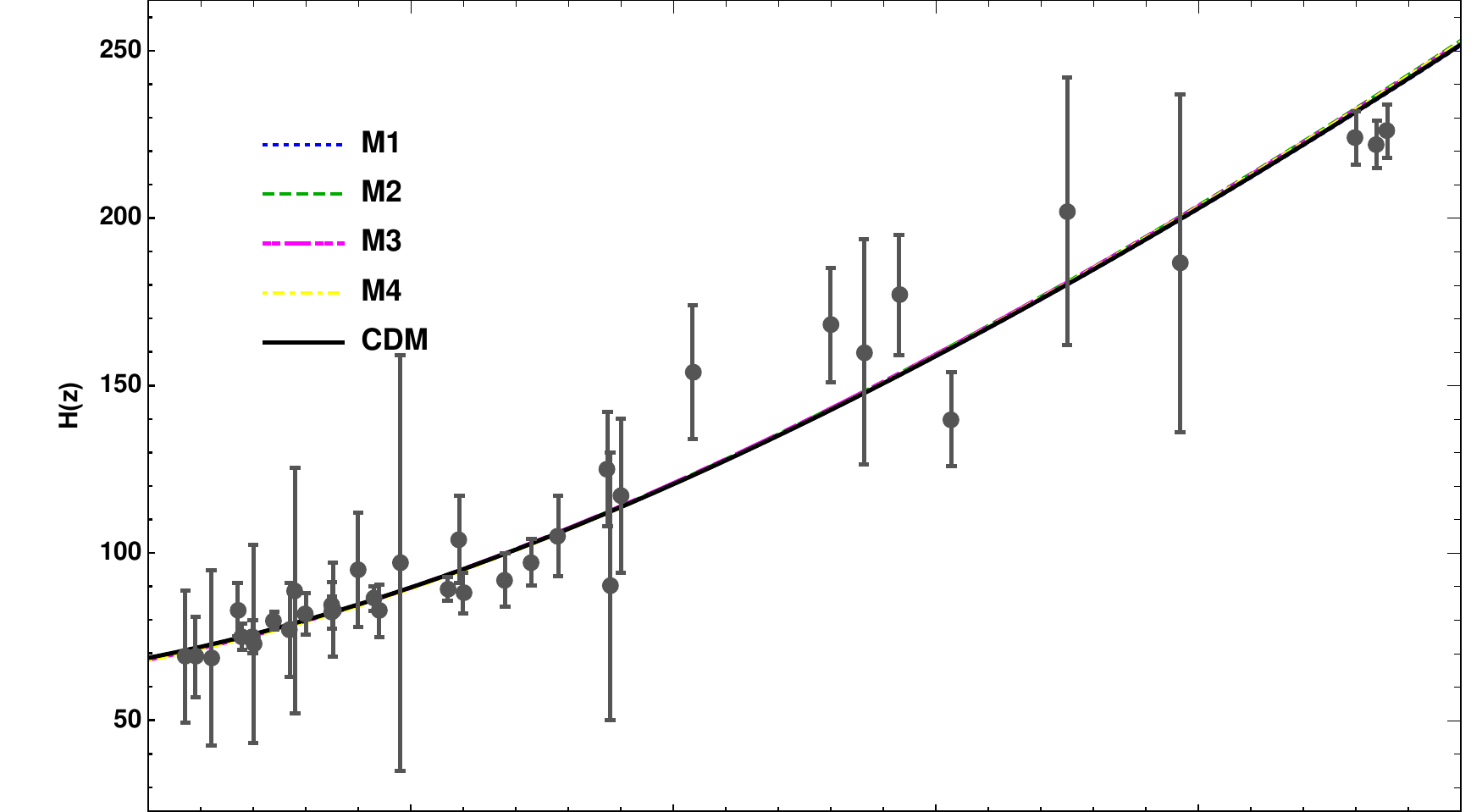}\\
	\includegraphics[width=0.5\linewidth]{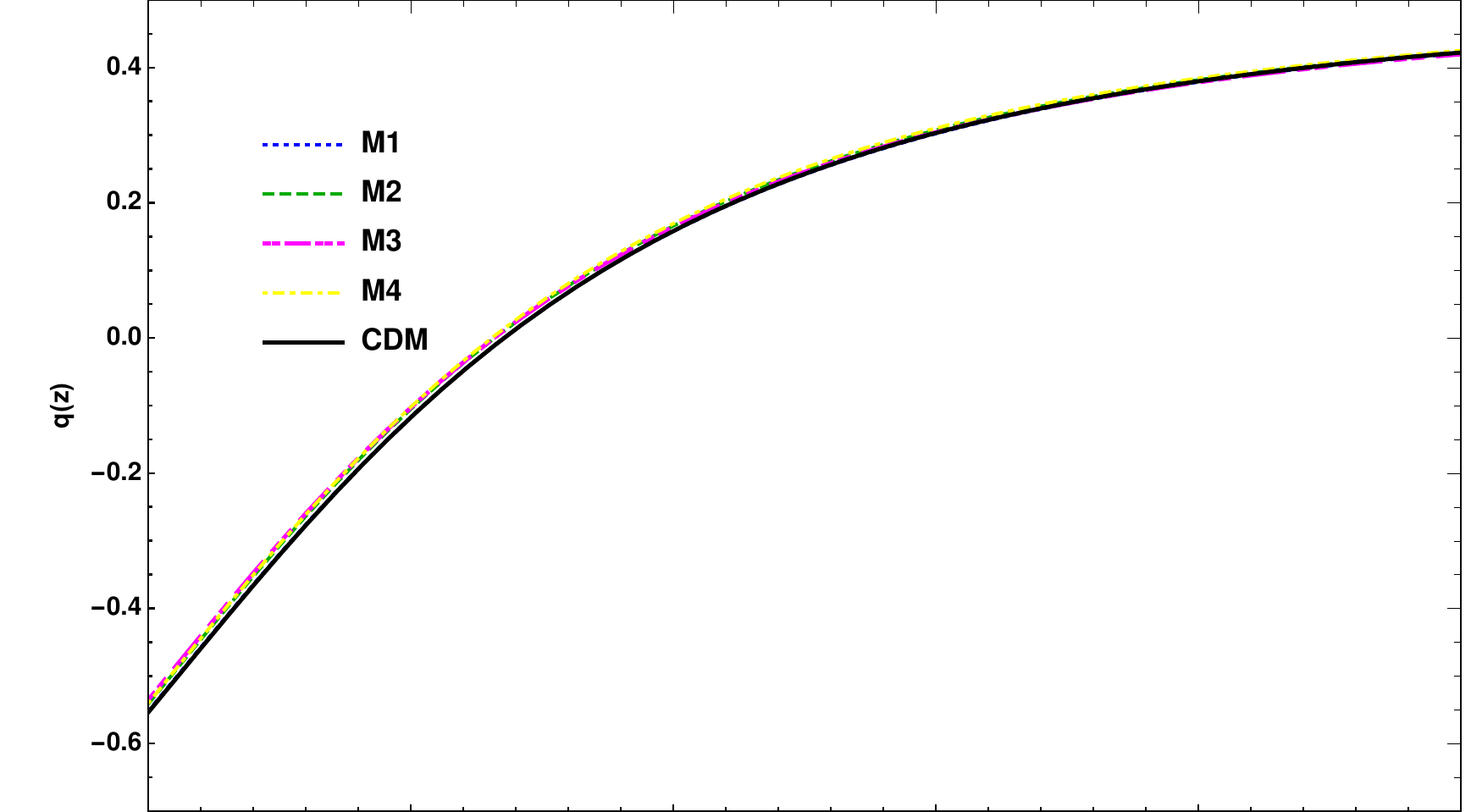}}
\gridline{		\includegraphics[width=0.5\linewidth]{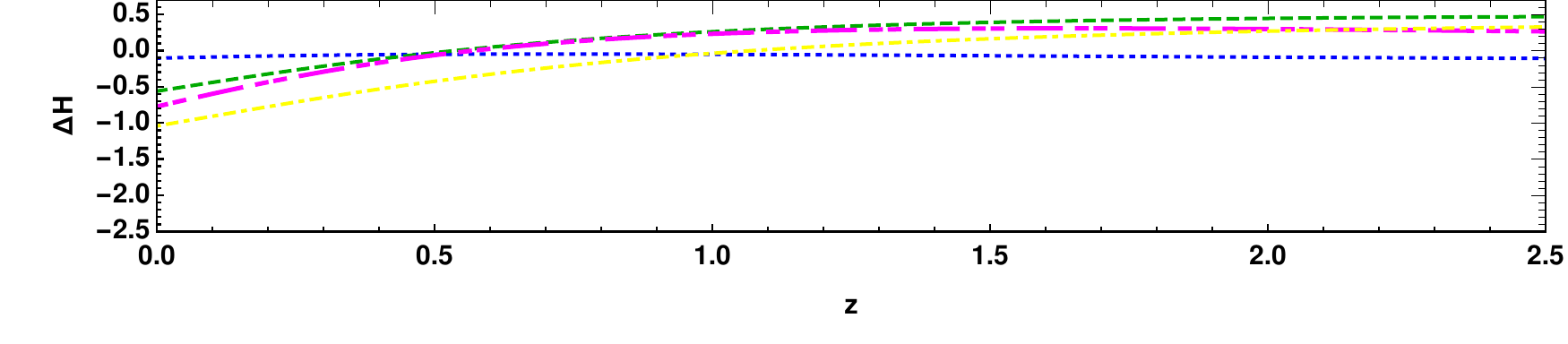}\\
	\includegraphics[width=0.5\linewidth]{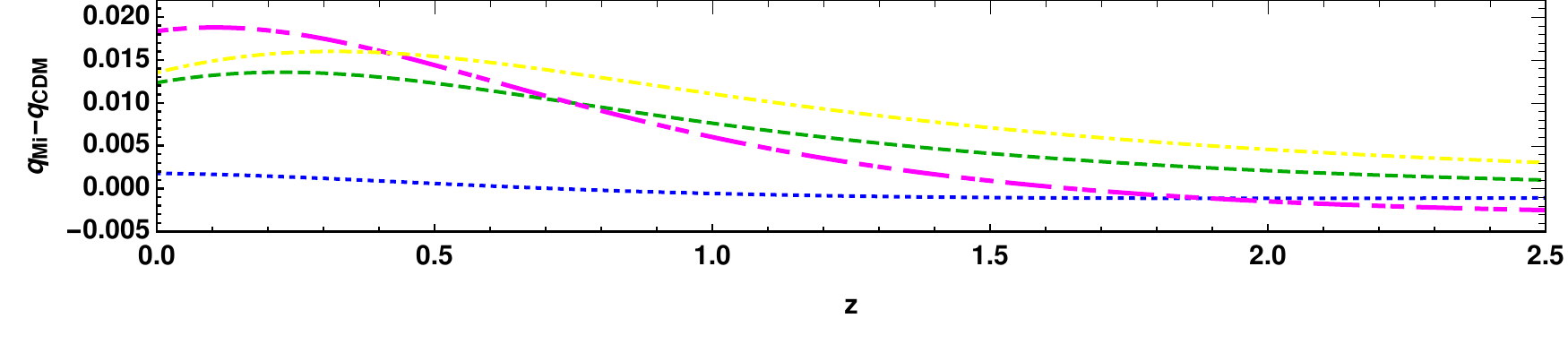}}
\caption{Left panel: theoretically predicted Hubble parameters and their relative differences as a function of redshift compared to the observational data from cosmic chronometers~\citep{Marra:2017pst}.Right panel: The redshift evolution of the deceleration parameter $q(z)$ and their differences as a function of redshift. 
For both panels, the values of the parameters obtained from Table~\ref{tab:best} with the total data has been used for the proposed viscous dark matter scenarios and the CDM model.	
}
\label{figh}
\end{figure}

To gain better insight into different models of viscous DM and compare it with the cold DM model, we will examine the evolution of the main cosmological quantities for the best values of the parameters obtained using total likelihood from  Table~\ref{tab:best} below.
As we know, one of the key cosmological parameters and a pivotal quantity in cosmology is the current value of the Hubble parameter since it is used to construct
the most basic time and distance cosmological scales. On the other hand, because of the impact of Hubble's expansion on the growth of matter perturbations, it is important to understand the behavior of $H(z)$ in various viscous DM cosmologies. Therefore, we plot the evolution of $H(z)$ in Figure~\ref{figh}. We present in the bottom panel of
Figure~\ref{figh}, the relative difference $\Delta H(z)= 100\times[H_{\rm Mi}-H_{\rm CDM}/H_{\rm CDM}]$ of the
Hubble parameters with respect to the CDM. As we see,  the quantity $\Delta H(z)$ is negative for all redshifts in the M1 case, but in other scenarios, is
positive (negative) for large (small)
redshifts.\\
As a supplementary information, we present another purely kinematical parameter, which is called the deceleration parameter, $q=-\frac{\ddot{a}}{a H^2}$ in the right panel of Figure~\ref{figh}.
$H(z)$ and $q(z)$ parameters are independent of any gravity theory, and both of them are only related to scale factor or redshift. We remind that parameter $ q > 0 $ corresponds to the deceleration phase of the universe expansion and $ q < 0 $ determines the accelerated phase of expansion. If we compare this panel with the $H(z)$ parameter,  we observe that the M1 case behaves similar to the CDM model and the transition redshift from deceleration to acceleration for both models occurs at $z_t\simeq 0.678$. We see at
early times for all models, $q$ tends to $ 1/2 $ as expected for the matter-dominated epoch. It should be noted that $q_{\rm Mi}-q_{\rm CDM}$ is more for the M3 model (that  for this case  $\xi \propto H^m$) in the present time ($z=0$).\\
\begin{figure}
	\begin{center}
		\includegraphics[width=10cm]{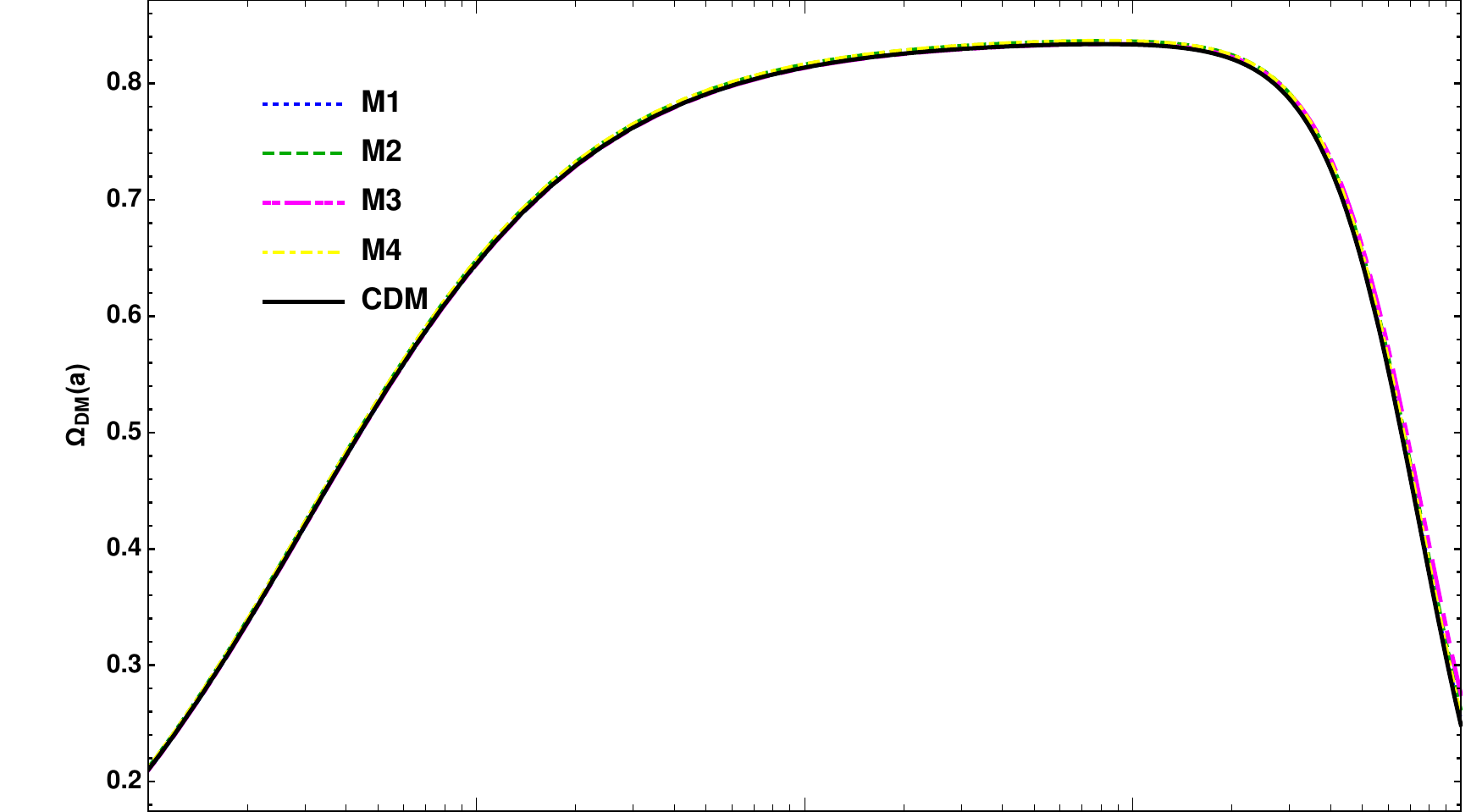}
		\includegraphics[width=10cm]{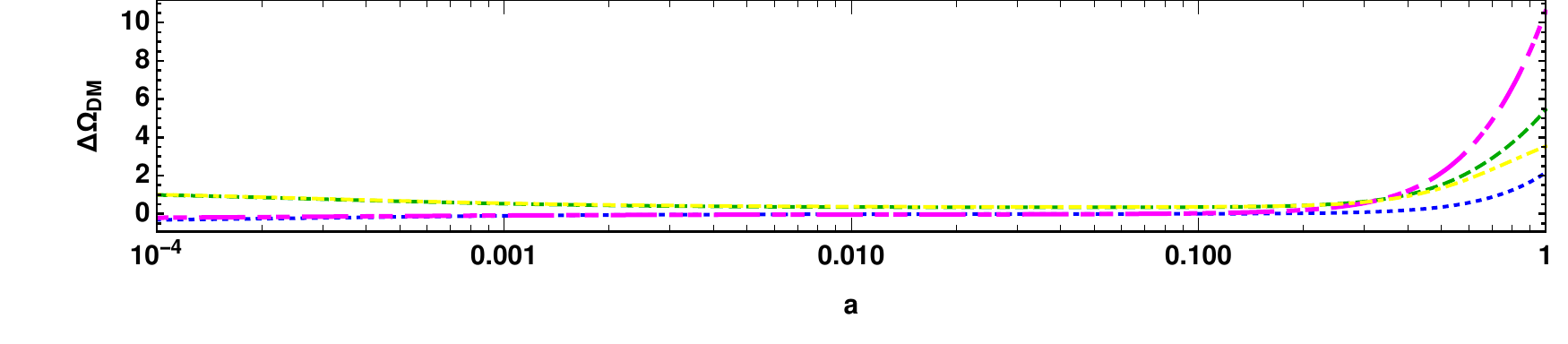}
		\caption{The evolution of the fractional energy density of viscous dark matter in terms of scale factor.
			We use the best fit values from Table~\ref{tab:best} for total data.
		}
		\label{figom}
	\end{center}
\end{figure}
We have plotted the evolution of the fractional energy densities of viscous dark matter in Figure~\ref{figom}. As we expect, all viscous DM models mimic the CDM scenario and the universe evolves from a radiation-dominated phase to a matter-dominated epoch, and later, as the energy of the dark matter decreases, the universe gets dominated by the dark energy in the recent universe. Given the results obtained from MCMC analysis, the density ratio $\Omega_{DM}/\Omega_B$ for M3  model is $\simeq 6$, which is larger than the CDM model ($\simeq 5$).\\
The last quantity that we plotted in this section is the equation of state parameter for viscous DM, $ w = \frac{P}{\rho}=\frac{-3 \tilde{\xi}(\rho_{\rm vdm},H,\rho_{\rm vdm}H)H}{\rho} $. As we see in Figure~\ref{figw}, in the past (when $a \rightarrow 0$), the viscous DM models behave like CDM ($ w\rightarrow 0 $), and currently, the value for $w$ is negative for all models. The most negative $w$ are respectively for the M2, M3, M4 and M1 models. As we see, the M1 model with the constant bulk viscosity coefficient is close to the CDM model for all scale factors ($ w_{{}_{\rm M1}(z)}=-0.016 $).
\begin{figure}
	\begin{center}
		\includegraphics[width=10cm]{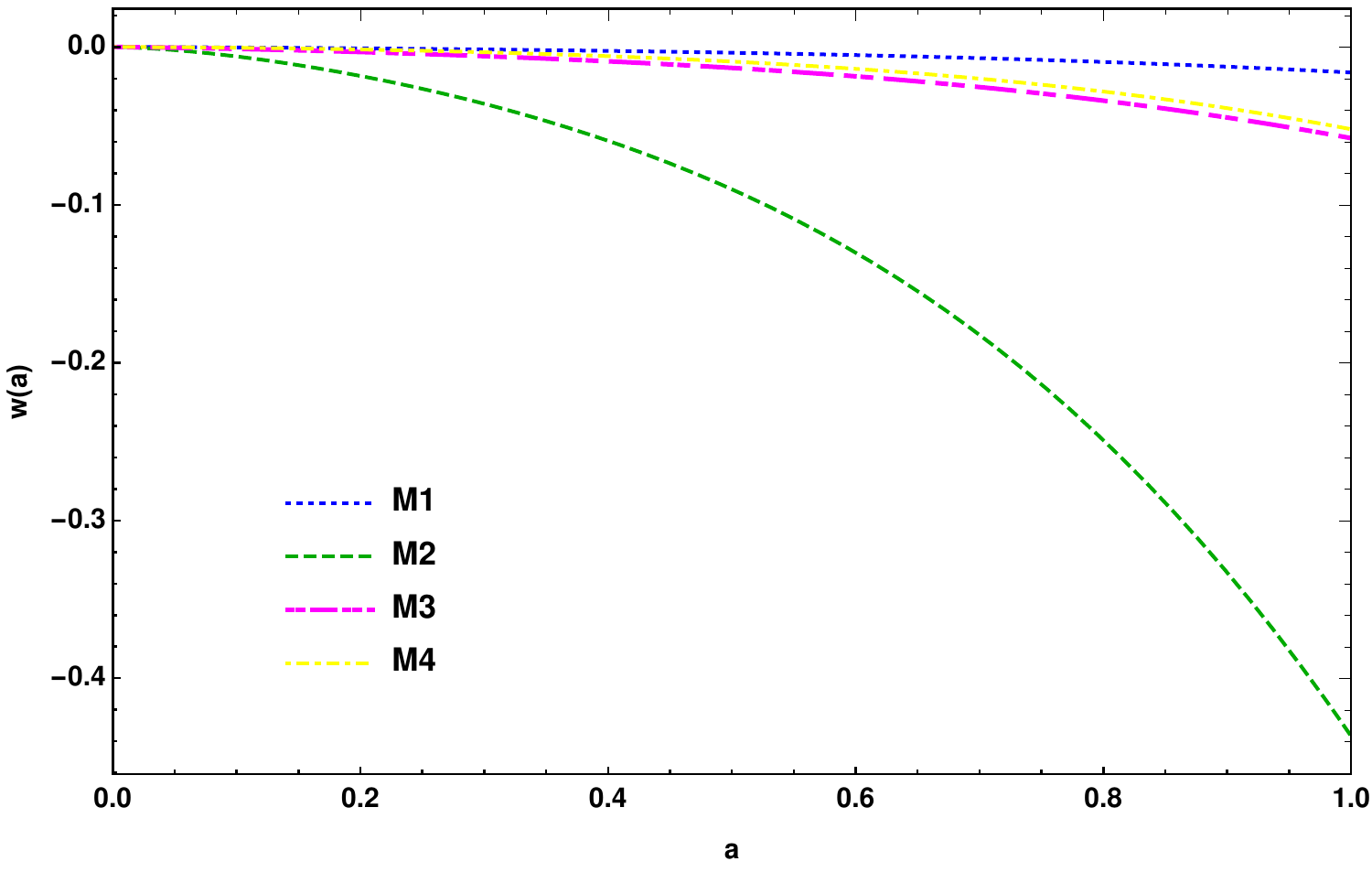}
		\caption{The evolution of the effective equation of state of viscous dark matter as a function of factor scale using use the best fit values from Table~\ref{tab:best} for total data.
		}
		\label{figw}
	\end{center}
\end{figure}
\subsection{Perturbation Level}\label{per}
\begin{figure}
	\gridline{	\includegraphics[width=0.5\linewidth]{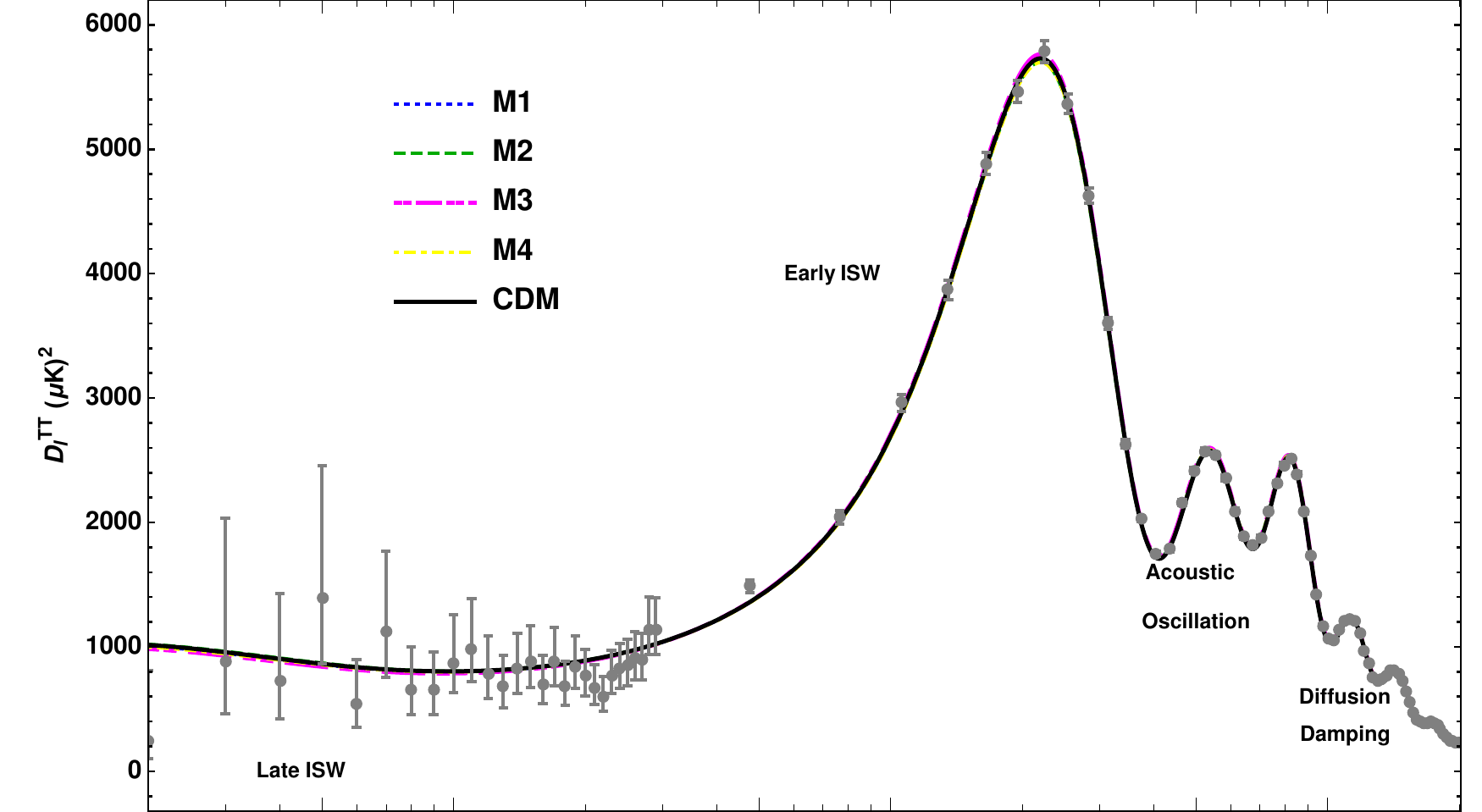}\\
		\includegraphics[width=0.5\linewidth]{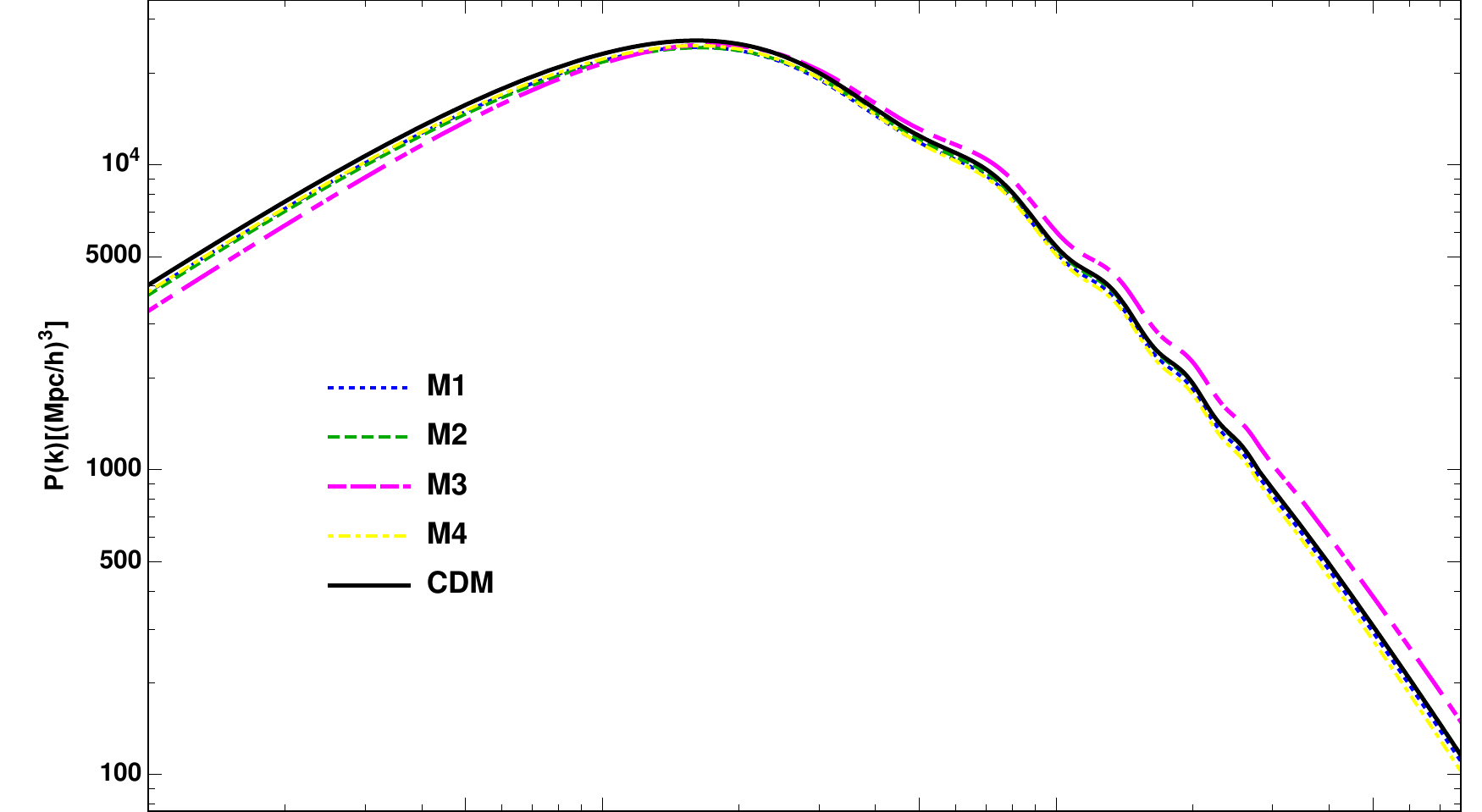}}
	\gridline{		\includegraphics[width=0.5\linewidth]{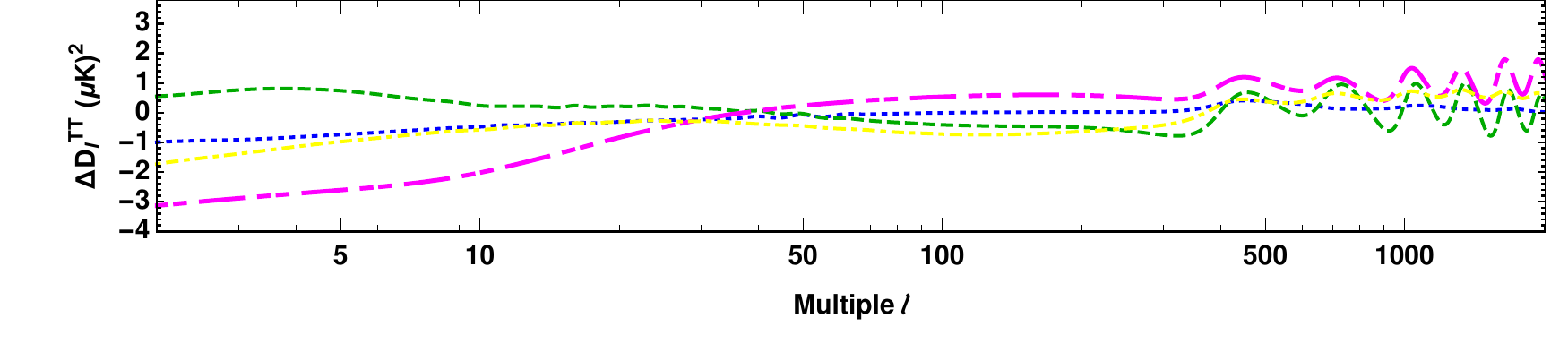}\\
		\includegraphics[width=0.5\linewidth]{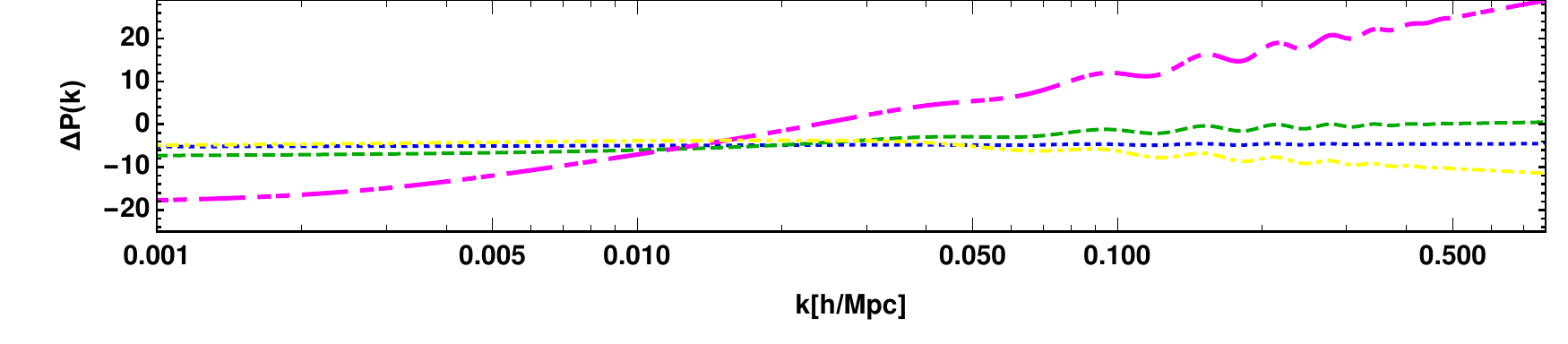}}
	\caption{Left panel: Temperature anisotropies in the CMB calculated for various viscous DM. The bottom part of the panel displays the relative temperature differences between the models and the CDM model. Physical processes which provide dominant contributions on different scales are quoted~\citep{Sugiyama:2014lga}. Right panel: Non-linear matter power spectrum in $z=0$ for various viscous DM and CDM models and the fractional difference between them. As we see, all viscous dark matter cases suppress $P(k)$ at large scales. We use the best-fit values of various free and derived parameters that have been shown in Table~\ref{tab:best} for total data.	
	}
	\label{figtt}
\end{figure}

\begin{figure}
	\gridline{	\includegraphics[width=0.5\linewidth]{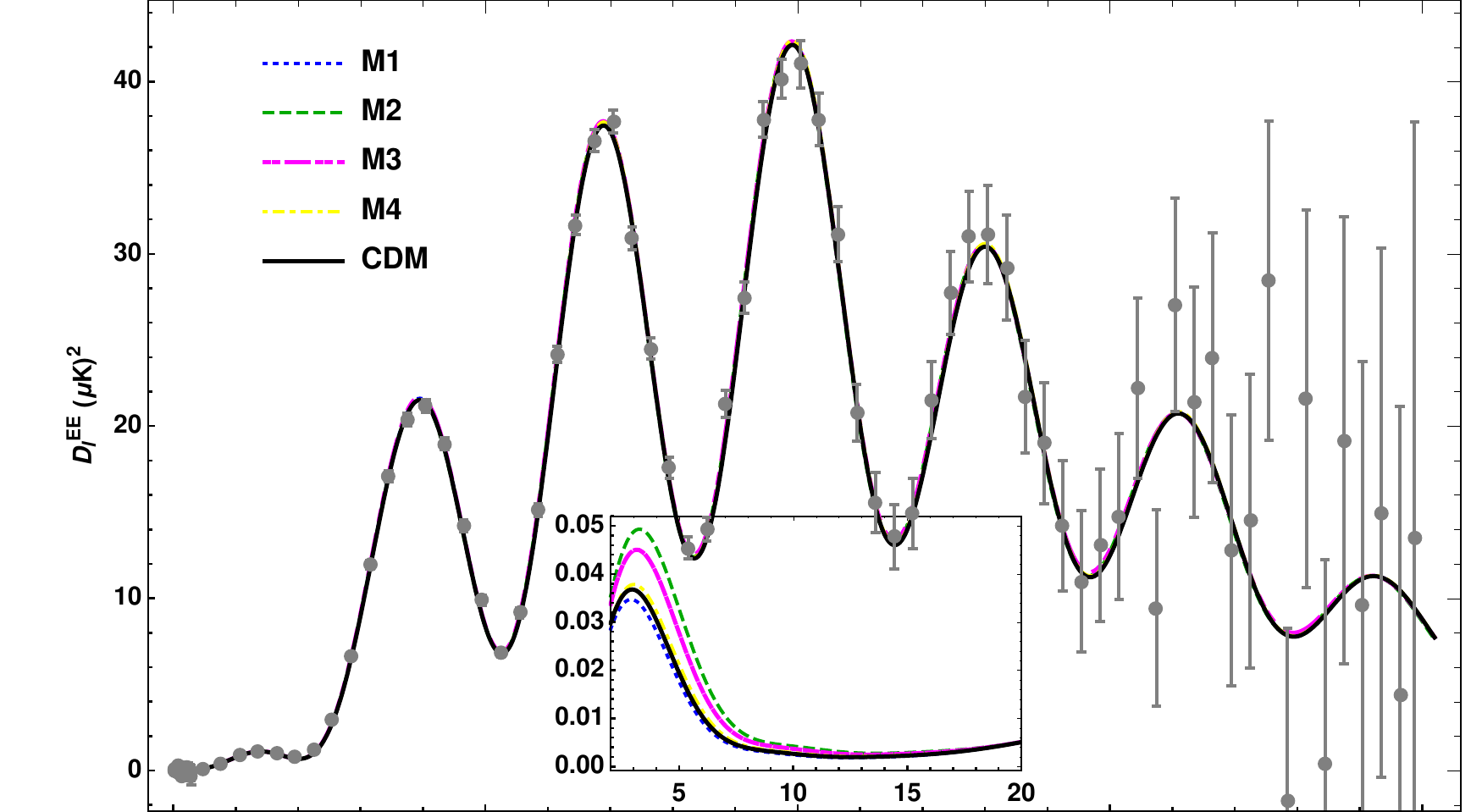}\\
		\includegraphics[width=0.5\linewidth]{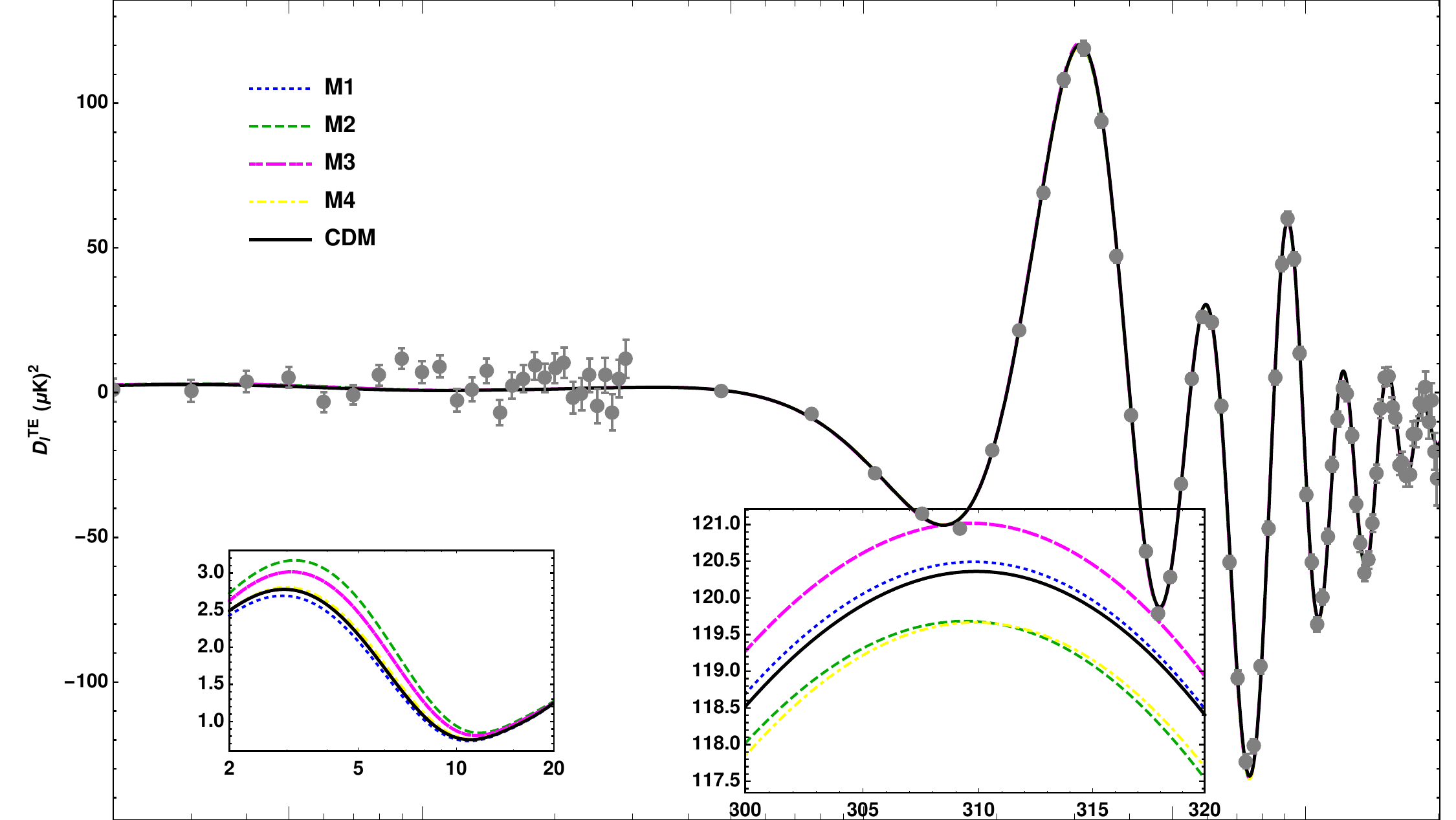}}
	\gridline{		\includegraphics[width=0.5\linewidth]{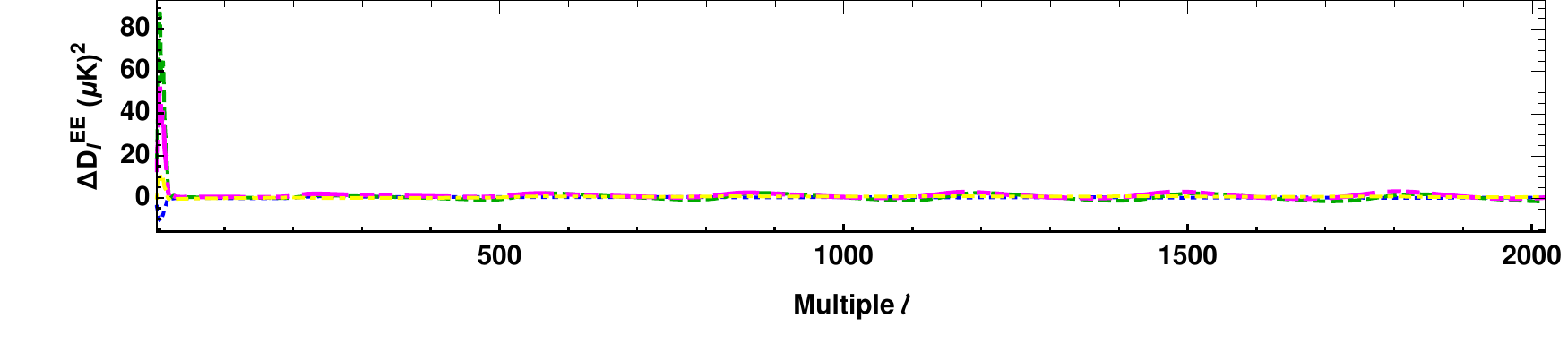}\\
		\includegraphics[width=0.5\linewidth]{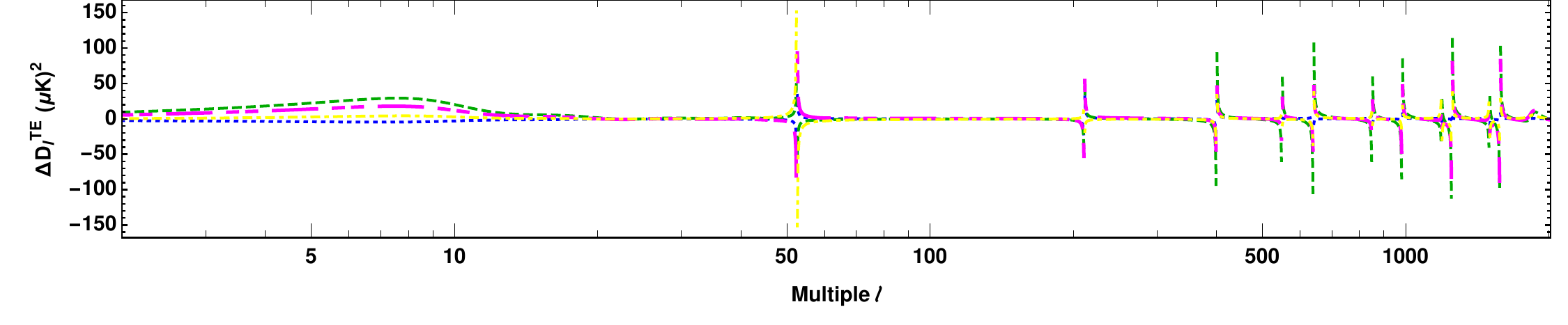}}
	\caption{Planck 2018 EE (left panel) and TE (right panel)power spectra using the best-fit values of various parameters have been obtained in Table~\ref{tab:best} for total data .	
	}
	\label{figee}
\end{figure}

\begin{figure}
	\gridline{\fig{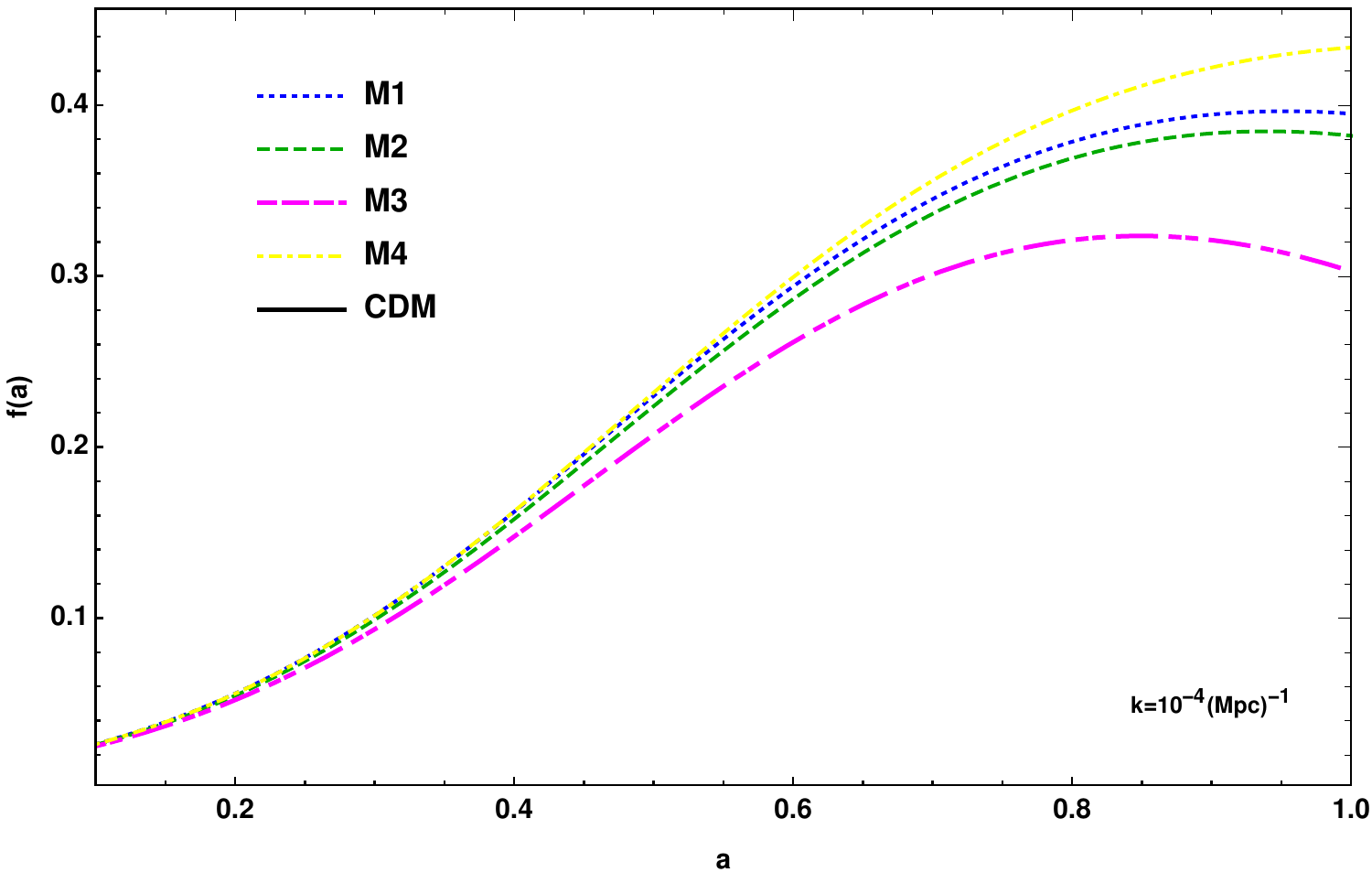}{0.5\textwidth}{}
	\fig{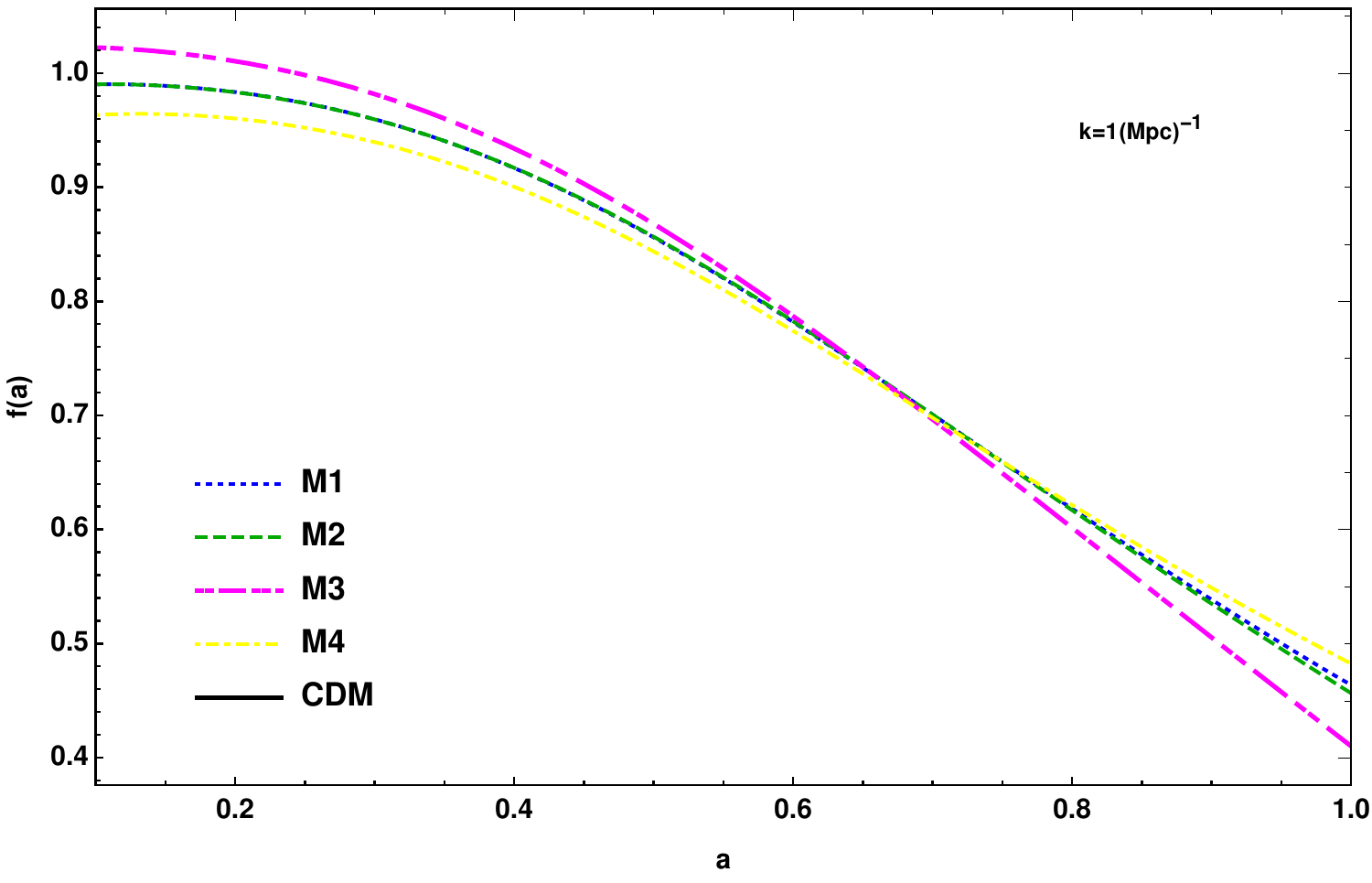}{0.5\textwidth}{}
}
\caption{	
	The linear growth rate, f (the scalar perturbations) for mode k = $10^{-4}$ (the galaxy
	cluster scale)and 1 Mpc$^{-1}$ (the dwarf galaxy scale).
}
\label{figf}
\end{figure}

\begin{figure}
	\begin{center}
		\includegraphics[width=10cm]{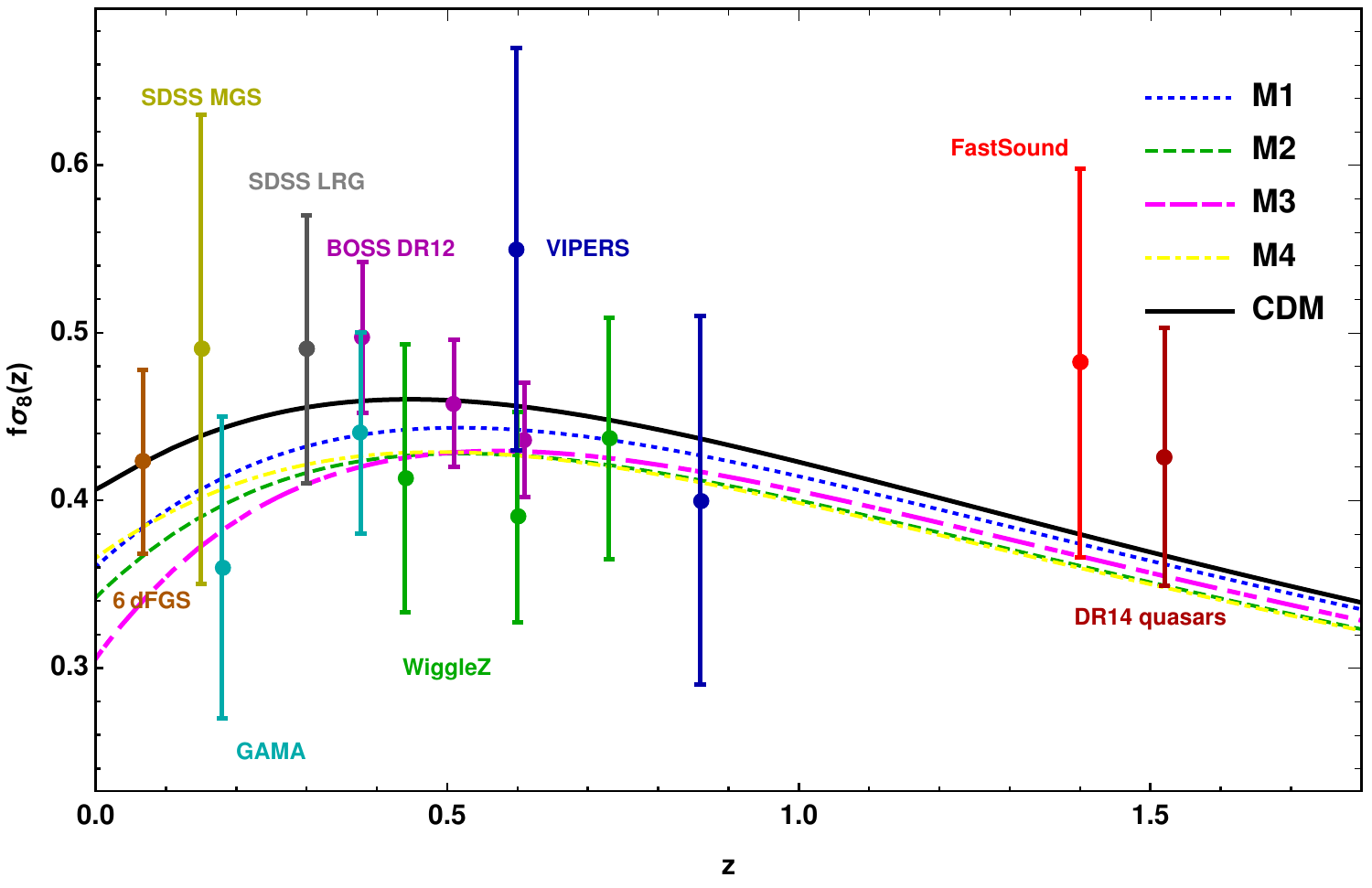}
		\caption{The growth rate of matter fluctuations for our baseline
			viscous DM models compared to $\Lambda$CDM
			model. The observational constraints
			are taken from \citep{Aghanim:2018eyx}.
		}
		\label{figfs8}
	\end{center}
\end{figure}

In this subsection, we turn to the perturbation spectra. To draw the plots, we have
used the best-fit values of the model parameters obtained
from all the observational datasets we have used in
this work. First, we display the power spectra of the anisotropies in the CMB temperature, calculated for viscous DM models, along with that for CDM in Figure~\ref{figtt}.
As we can see in the lower multipoles (for $l<50$), there is a substantial deviation of $\Delta D_l^{\rm TT}$ of the viscous DM models from that of the CDM scenario. For higher multipoles (for $l > 50$), we do not observe any significant differences in the curves.
We can explain this point by noting that the CMB temperature is determined only by the energy density component of the energy-momentum tensor, $T_{00}$, whereas the density perturbations also depend upon the diagonal elements, $T_{\rm ii}$, which include the pressure. Since the viscous DM scenarios own pressure, this causes a difference between the viscous DM models and CDM. 
Hence, as the evolution of density perturbations unfolds, small-scale structures are clearly suppressed in viscous DM compared to CDM, which can be seen in the right panel of Figure~\ref{figtt}. The amount of suppression is larger at smaller $k$'s. Respectively, models M4, M1, and M2 suppress the nonlinear matter power spectrum for $ k>0.01 $, but the M3 model, which has a viscosity coefficient proportional to the Hubble parameter, enhances the evolution of density perturbations  compared to the CDM model at such scales.
\\
As we can see in the left panel of Figure~\ref{figtt} and panels of Figure~\ref{figee}, the position of the acoustic peaks has not shifted that much, which means that the viscous dark matter does not have much effect on the CMB scales.
We also see a difference between models in the panels of Figure~\ref{figee} for low $\textit{l}$. This is due to the integrated Sachs-Wolfe (ISW) effect on such scales.
Also, the small $l$'s of  CMB (TT and, even better, EE) give information on the reionization history. We obtain the redshift of the reionization using the best values of parameters to be 7.23, 8.40, 8.05, and 7.57 respectively. for models M1, M2, M3 and M4. M2 and M3 have the biggest differences from the standard model ($z_{\rm reio,CDM}=7.78$).\\  
The present galaxy surveys provide powerful observational data for the combination $f\sigma_8(z)$, where $f$ is the linear growth rate of the density contrast, which is related to the
peculiar velocity in the linear theory, being defined by $f(a)=a \frac{\delta'(a)}{\delta(a)}$. On the other hand, $\sigma_8(a)=\sigma_8(a=1)\frac{\delta(a)}{\delta(a=1)}$ is the root-mean-square mass fluctuation in spheres with the radius $8h^{-1}$ Mpc. Here, we calculate the total density contrast of the matter (baryons + dark matter) as
\begin{equation}
\delta(a)=\frac{\delta_b \rho_b+\delta_{\rm dm} \rho_{\rm dm}}{\rho_b+\rho_{\rm dm}}.
\end{equation}
The linear growth rate, $f$, is an excellent tool
for distinguishing between various dark matter theories
based on the growth of large-scale structures.
In Figure~\ref{figf} we show the linear
growth factor $f(a)$ for the standard models and different viscous DM scenarios studied in this work for the best-fit values for modes $k=10^{-4}$Mpc$ ^{-1} $ (corresponding to galaxy cluster scale) and $k=1$Mpc$^{-1}$. Since the linear growth rate is independent of the wave number $k$ at the large scale factors (the low redshift regime), we see that for $a\rightarrow 1$, two modes behave similarly.  The bottom panel shows that for small-scale factors the growth function goes to unity, which corresponds to the
matter-dominated Universe, and the amplitude of matter perturbations decreases at low redshifts in all scenarios with the dominance of dark energy. We observe that the suppression of the amplitude of matter fluctuations in M3 and M4 models that are proportional to the Hubble parameter starts sooner in comparison with the other scenarios.
The change in $\sigma_8$ in the viscous DM scenarios can be seen as the relative decrease in $P(k)$ in the right panel of Figure~\ref{figtt} (although $\sigma_8$ is computed from the linear rather than non-linear power spectrum), thus improving the tension between these CMB-derived predictions and the actual LSS data. Looking at Figure~\ref{figfs8}, we notice that considering viscosity has the influence of slowing down the evolution rate of the dark matter perturbations. This means that structures cluster slower in the viscous dark matter models as we predicted from the right panel of  Figure~\ref{figtt}. In particular, scenarios M4 and M3 suppress the power spectrum more in comparison with the other CDM scenario. This result is in agreement with~\cite{Barbosa:2017ojt}. They showed that  viscosity can suppress the growth of perturbations and delay the formation of nonlinear regime.
\begin{figure}
	\begin{center}
		\includegraphics[width=8cm]{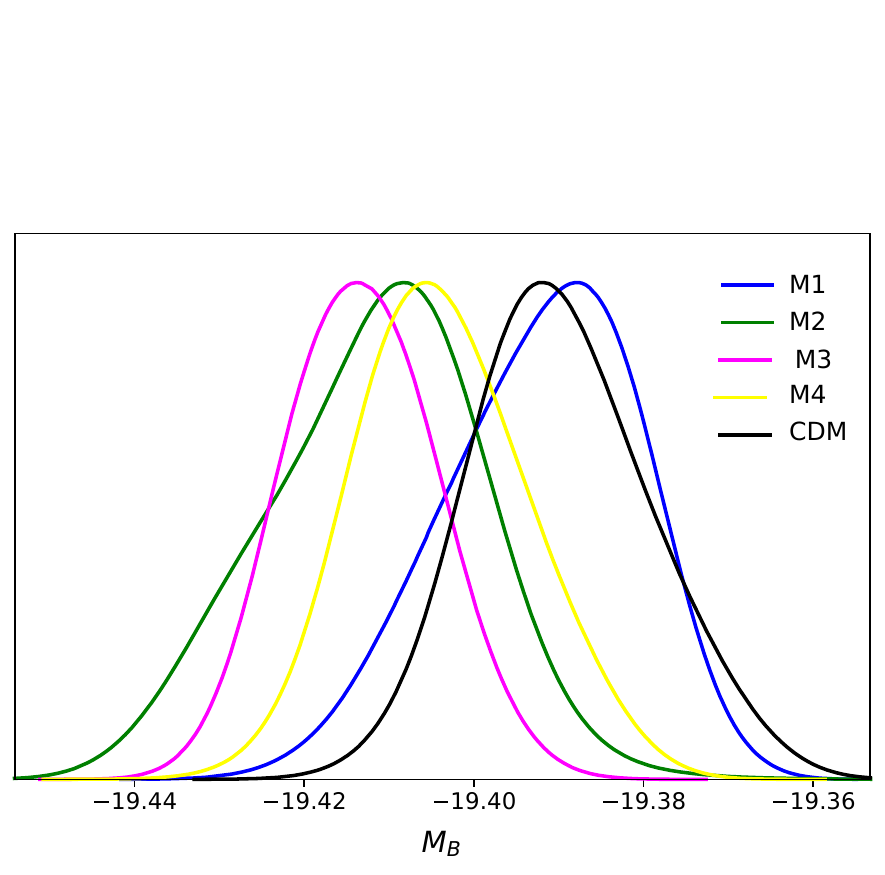}
		\caption{Comparison between the absolute magnitude, $ M_B $ for all considered models in this study.
		}
		\label{figmb}
	\end{center}
\end{figure}
\section{Large-scale cosmic tensions}\label{sec4}
In this section, we reported the obtained results for the $H_0$ and $S_8$ tensions.\\
\subsection{$H_0$ Tension}
The Hubble tension is one of the most lingering tensions in cosmology. This tension emerged
with the first release of Planck results and has grown in significance in the past few years. The last value of $ H_0 $ tension is obtained from the most recent analysis of Planck 2018 for the early-universe observation $H_0 = 67.44\pm0.58$ km$s^{-1}$Mp$c^{-1}$ for the $\Lambda$CDM model~\citep{Efstathiou:2019mdh}. While the late-universe observation of the Cepheid-calibrated SNe Ia distance ladder by the SH0ES team gives the result of $H_0 = 74.03\pm1.42$ km$s^{-1}$Mp$c^{-1}$~\citep{Riess:2019cxk}.
The big discrepancy between the two $H_0$ values  reaches $\sim4\sigma $~\citep{Verde:2019ivm,Freedman:2019jwv,DiValentino:2020zio}.
We wondered if the tension could be relieved by manipulating the properties of dark matter.
As the results of  Table~\ref{tab:best} and Figure~\ref{fig1} demonstrate, none of the models of viscous dark matter diminishes the $ H_0 $ tension.
On the other hand, $H_0$ tension is equivalent to the mismatch of the Pantheon SNIa absolute magnitudes, which when calibrated using the CMB sound horizon and propagated via BAO measurements to low $z$ have a value $M_B=-19.401 \pm 0.027$ mag~\citep{Camarena:2019rmj}, while when calibrated using the Cepheid stars have a value $M_B=-19.244 \pm 0.037$~\citep{Camarena:2019moy}. Given figure~\ref{figmb}, for all models, the absolute magnitudes obtained is $\sim -19.41$, which is consistent with the CMB observations. Therefore, none of these models really help us in addressing the $H_0$ tension.

\subsection{$S_8$ Tension}\label{sec:s8}

Cosmic structures were formed based on primordial density fluctuations that emerged from the inflation era due to gravitational instability.
By studying the large-scale structure of the Universe and its evolution in cosmic epochs, it is possible to trace the growth of cosmic perturbations. The main difference between the standard model and other dissipative approaches lies in the behavior of the cosmological perturbations. While perturbations in standard cosmology are always adiabatic, dissipative models of the dark sector are intrinsically non-adiabatic. Associated with this observable, one of the current cosmological tensions is growth tension. It has come about as the result of the discrepancy between the Planck value of the cosmological parameters $\Omega_m$ and $\sigma_8$ from weak gravitational lensing surveys, cluster counts, and redshift space distortion data. Since these two parameters have degenerate effects in the lensing surveys, usually $S_8=\sigma_8\sqrt{\Omega_m/0.3}$ (weighted amplitude of matter fluctuations) is used as a parameter to compare the consistency with other observations.
Measurements of the
CMB temperature and polarization anisotropies by Planck yield $ S_8 = 0.834\pm0.016 $. On the other hand, weak gravitational lensing surveys provide constraints
via cosmic shear, e.g. $ S_8 = 0.759^{+0.024}_
{-0.021} $ from the Kilo-Degree Survey (KiDS-1000)~\citep{KiDS:2020suj} and
$ S_8 = 0.780^{+0.030}_{-0.033} $ from Hyper Suprime-Cam (HSC)~\citep{HSC:2018mrq}. The combination of shear and galaxy clustering from the three-year data of DES yields $S_8 = 0.776^{+0.017}_{-0.017 }$ and a combination of shear, clustering, and galaxy
abundance from KiDS-1000 $S_8 = 0.773^{+0.028}_
{-0.030}$, while galaxy cluster counts from SPT-SZ
report $ S_8 = 0.766 \pm0.025 $ and eROSITA results favor $ S_8 = 0.791^{+0.028}_{-0.031} $.
We adopt the following estimator in order to quantify the discordance or tension in current determinations
\begin{figure}
	\begin{center}
		\includegraphics[width=7cm]{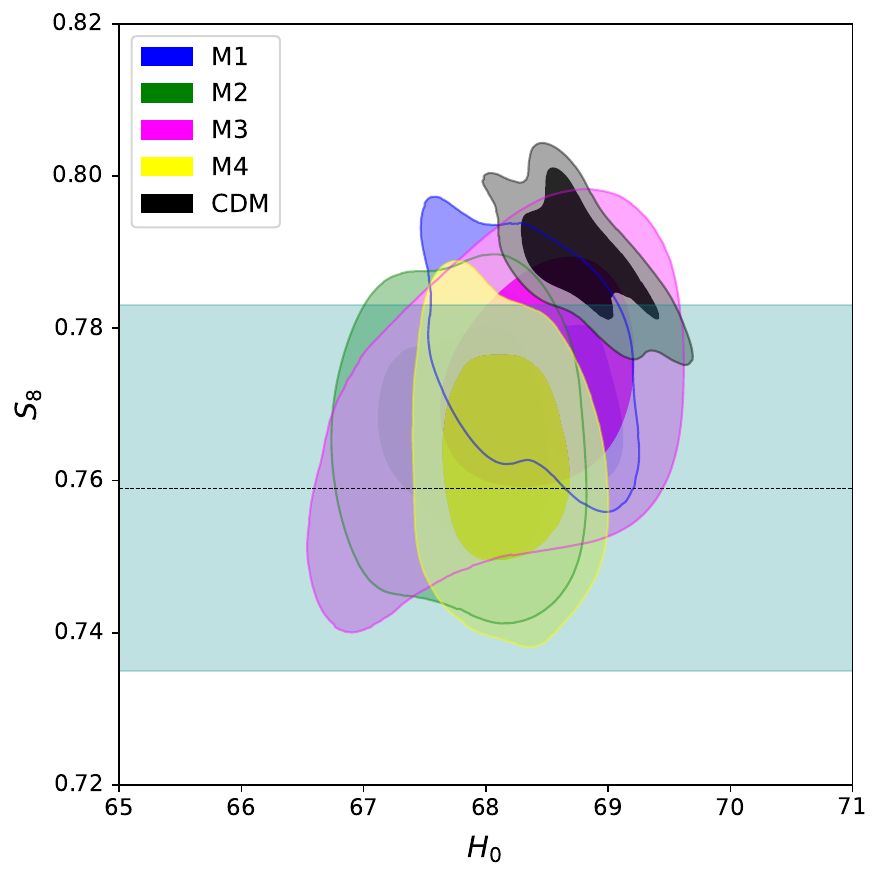}
		\caption{The preferred ranges of the parameters $S_8 \equiv\sigma_8 (\Omega_m /0.3)^{0.5}$ and $ H_0 $ for all reviewed models using total data. For reference, the measurements by $S_8=0.759\pm0.02$ by KiDS-1000 are also shown.
		}
		\label{figs8h}
	\end{center}
\end{figure}
\begin{equation}
T_{S_8}=\frac{|S_8^i-S_8|}{\sqrt{\sigma^2_i+\sigma^2}}.
\end{equation}
Considering the reported value of the Kilo-Degree Survey KiDS-1000 as observational data, as we see in Table~\ref{tab:s8}, all models of viscous dark matter are able to decrease this tension. More than any of the viscous models, M4  can reduce the tension between Planck and KiDs-1000 data, respectively.
\begin{table}
	\centering
	\caption{The result of $T_{S_8}$  for all reviewed models in this study considering the reported value of  KiDS-1000 as observational data.}
	\begin{tabular}{|l|c|c|c|c|c|}
		\hline
		Model&M1&M2&M3&M4&CDM\\
		\hline
		$T_{S_8}$&$0.58\sigma$&$0.23\sigma$&$0.50\sigma$&$0.15\sigma$&$1.25\sigma$\\
		\hline
	\end{tabular}\label{tab:s8}
\end{table}
\section{Bounds on Neutrino Mass in k-Dependent dark matter}\label{mass}
\begin{figure}
	\centering
	\includegraphics[width=12cm]{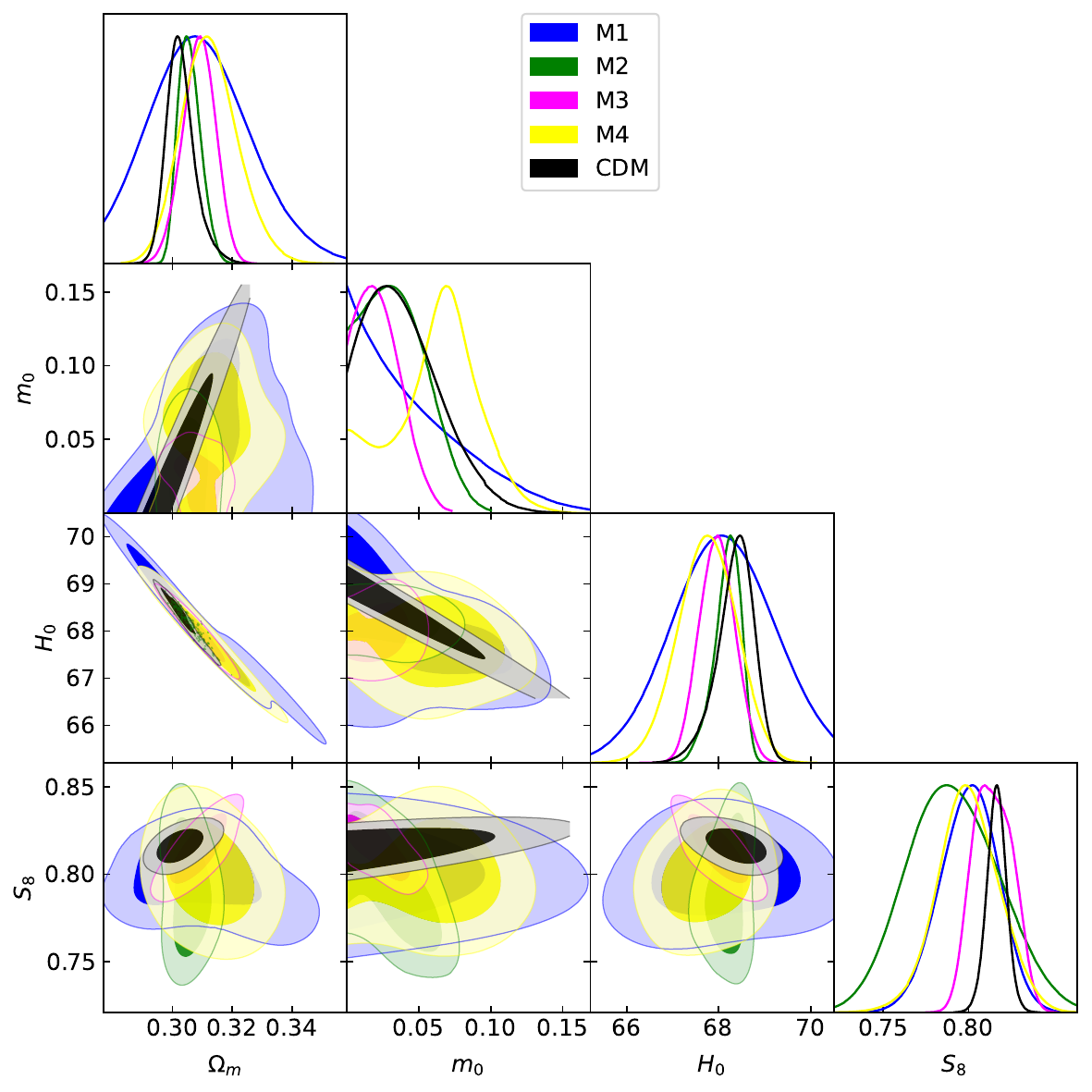}
	
	\caption{68\% and 95\% constraint contours on the matter density parameter $ \Omega_m $ , the lowest neutrino mass $ m_0 $, the Hubble parameter, $H_0$ and $S_8$. It is obtained by considering the Planck+BAO+CC+SNIa+LSS data
		\label{fig11}}
\end{figure}

In the previous section, we obtained that all models of viscous dark matter are able to reduce the $S_8$ tension. On the other hand, we know that the inclusion of massive neutrinos in cosmology could also suppress the matter power spectrum on the small length scales, similar to how the viscosity does it~\citep{Battye:2013xqa, Anand:2017ktp}.
Massive neutrinos have this important property that they are relativistic in the early universe and contribute to the density of radiation and,  in the late time when they become non-relativistic, they contribute to the density of the total matter. After they become non-relativistic, due to the collisionless nature of neutrinos, they are allowed to free-stream on scales $k > k_{\rm fs} $, where  $ k_{\rm fs} $ is the wavenumber corresponding to the scale of neutrino free-streaming. Hence they will wash out the perturbations on length scales smaller than the characteristic scale $ k_{\rm fs} $. This leads to the suppression of power on small scales in the matter power spectrum.\\
An open issue in particle physics is the determination of the absolute mass of neutrinos. Due to the role of neutrinos in the evolution of the background as well as the formation of structures in the universe, apart from particle physics experiments, cosmological observations can also be used to extract this information about neutrinos.
In~\cite{Battye:2013xqa}, it was argued that the mismatch between the Planck CMB observations
and LSS surveys could be a signature of nonzero neutrino masses and hence they obtained constraints on the sum of masses of neutrinos. Also in \cite{Anand:2017ktp}, where constant effective viscosity was assumed, an upper bound on the sum of the neutrino masses was found. We therefore decide to constrain the cosmological bound on neutrino masses by including various time-varying viscous DM models, elaborated in the previous sections. We follow \cite{Ma:1995ey} to describe the perturbation equations for massive neutrinos using publicly available \texttt{CLASS} code. \\
For this purpose, we repeated the MCMC analysis using \texttt{MONTEPYTHON-v3} by considering the Planck+BAO+CC+SNIa+LSS data to add the mass of the lightest neutrino mass as a free parameter ($ m_0 $).  In Figure~\ref{fig11} and Table~\ref{tab:m0}, we
show the maximum allowed values of $ m_0 $ at 2-$ \sigma $ level using the joint MCMC analyses. We found that the inclusion of the time-varying viscous model tightens up the bound on $ m_0 $ for all viscous DM models. 
\section{Discussion and Conclusion}\label{con}
With the aim of finding a model more compatible with the universe and reducing the current controversial tensions in cosmology, this study questioned one of the properties of dark matter in the Standard Model of cosmology. We include a physical dissipation mechanism in the description of dark matter such as viscosity. In addition to the constant bulk viscosity case which was studied previously by \cite{Anand:2017ktp}, we assessed the cases in which the bulk viscosity has a dependence on $\rho_{vdm}$, the Hubble parameter, or both. We noticed that with such dependences on viscosity, the $\chi^2$ for the cosmological datasets is reduced. In a more quantitative term, $\chi^2$ of the models satisfy $M4<M3<M2<M1$. Noting that we have added more parameters, we computed the AIC criterion for these models and noticed that the same hierarchy also appears for the models, namely AIC values for the models are ranked as $M4<M3<M2<M1$.   Our results in Tables~\ref{tab:best} and Figure~\ref{figs8h} show that the assumption of viscous dark matter can only reduce the $S_8$ tension, and not the $H_0$ tension. In order to relax the observed tensions $H_0 $ and $ S_8 $ simultaneously, perhaps in addition to viscous dark matter, the presence of relativistic components or early dark energy, should also be considered. The LSS-Planck tension has been advocated many times in the past as a signature of massive neutrinos in cosmology. Figure~\ref{fig11} shows that in different scenarios of viscous dark matter, we considered in this paper, massive neutrinos become more constrained than the cold dark matter scenario.
\begin{table*}
	\centering
	\caption{2-$\sigma$ upper bounds on  the neutrino masses.}
	\begin{tabular}{|l|c|c|c|c|c|}
		\hline
		Model&M1&M2&M3&M4&CDM\\
		\hline
		$m_0$(ev)&$0.148$&$0.076$&$0.061$&$0.122$&$0.152$\\
		\hline
	\end{tabular}\label{tab:m0}
\end{table*}
\section{ACKNOWLEDGMENTS}
This work is supported by Iran Science Elites Federation Grant No M401543. This project also received funding support from the European Union's Horizon 2020 research and innovation programme under the Marie Sk\l{}odowska -Curie grant agreement No 860881-HIDDeN.
\section{DATA AVAILABILITY}
No new data were generated or analyzed in support of this research.
\bibliography{sample631}{}
\bibliographystyle{aasjournal}

\end{document}